\documentclass[11pt]{article}
\usepackage{amsmath}   
\usepackage{amsfonts}    
\usepackage{mathrsfs} 
\usepackage{xfrac} 
\usepackage{tabu}
\usepackage{color}

\usepackage{tikz}
\usetikzlibrary{trees,snakes}
\usepackage{verbatim}

\usepackage{amsmath}
\usepackage{amsfonts}
\usepackage{amssymb}
\usepackage{graphicx}
\usepackage{tikz}
\usetikzlibrary{calc}
\usepackage{pgfplots}
\usepackage{cleveref}
\usepackage{graphicx}
\usepackage{subcaption}
\usetikzlibrary{trees,snakes}
\usepackage{verbatim}

\usepackage{mathrsfs} 
\usepackage{xfrac} 
\usepackage{tabu}
\usepackage{color}
\usepackage{tikz}
\usepackage{bbm}

\definecolor{myred}{RGB}{255, 0, 0}
\definecolor{myblue}{RGB}{0, 0, 255}
\newtheorem{theorem}{Theorem}

\newtheorem{proposition}{Proposition}
\newtheorem{fact}{Fact}

\newcommand {\nn} {\nonumber}
\newcommand {\pr} {\mathbb{P}}
\newcommand{\IND}{\mathbbm{1}}

\newcommand{\dfn}{\stackrel{\triangle}{=}}

\newcommand {\bx} {\boldsymbol{x}}
\newcommand {\by} {\boldsymbol{y}}

\newcommand {\bX} {\boldsymbol{X}}
\newcommand{\calA}{{\cal A}}
\newcommand{\calB}{{\cal B}}
\newcommand{\calC}{{\cal C}}

\newcommand{\calF}{{\cal F}}

\newcommand{\calM}{{\cal M}}
\newcommand{\calN}{{\cal N}}

\newcommand{\calX}{{\cal X}}

\begin{document}
\thispagestyle{empty}
\title{Simple Majority Consensus in Networks with Unreliable Communication\\}
\author{\\ Ran Tamir (Averbuch), Ariel Livshits, and Yonatan Shadmi\\}
\maketitle
\begin{center}
The Andrew \& Erna Viterbi Faculty of Electrical Engineering \\
Technion - Israel Institute of Technology \\
Technion City, Haifa 3200003, ISRAEL 
\end{center}
\vspace{1.5\baselineskip}
\setlength{\baselineskip}{1.5\baselineskip}

\begin{abstract}
In this work, we analyze the performance of a simple majority-rule protocol solving a fundamental coordination problem in distributed systems - \emph{binary majority consensus}, in the presence of probabilistic message loss. Using probabilistic analysis for a large scale, fully-connected, network of $2n$ agents, we prove that the Simple Majority Protocol (SMP) reaches consensus in only three communication rounds with probability approaching $1$ as $n$ grows to infinity. 
Moreover, if the difference between the numbers of agents that hold different opinions grows at a rate of $\sqrt{n}$, then the SMP with only two communication rounds attains consensus on the majority opinion of the network, and if this difference grows faster than $\sqrt{n}$, then the SMP reaches consensus on the majority opinion of the network in a single round, with probability converging to $1$ exponentially fast as $n \rightarrow \infty$.   
We also provide some converse results, showing that these requirements are not only sufficient, but also necessary. \\

\end{abstract}

\clearpage
\section{Introduction}

The digital age has driven forth the need for easy and fast access to information. The world wide web has facilitated the existence of many useful multi-agent systems from messaging apps, to cryptocurrency \cite{vujivcic2018blockchain} and distributed data storage (or cloud services) \cite{yang2016construction, dingledine2001free}.  
However, the design of multi-agent systems inherently requires agents to communicate and coordinate according to a prescribed shared protocol in order to achieve a common goal. For example, messaging apps must always show messages in the same order to all participants in a conversation, which is challenging when user clocks are not necessarily synchronized \cite{fidge1987timestamps,mattern1988virtual}. Cryptocurrencies employ decentralized data structures to register currency transactions, which require a vast majority of users to agree upon its current state \cite{waldo2019hitchhiker}. Distributed data storage services must show consistent views of stored files in the presence of multiple concurrent reading and writing operations \cite{Liu2014consistency, kraska2009consistency}.

In the pursuit of developing such distributed protocols, much of the literature routinely makes two powerful assumptions. The first is that communication links are reliable \cite{chandra1996unreliable,hurfin1999simple,schiper1997early}, i.e., all messages between agents are eventually delivered. The second is that there exists an upper bound on the transmission delay of messages from one agent to another (usually the maximum propagation time of links) \cite{aguilera2010stumbling}. Nonetheless, communication networks are notoriously unreliable \cite{borran2007consensus, zielinski2007indirect, guerraoui2000consensus}. In fact, actual communication links may suffer from sudden crashes, resulting in messages in transit to be lost forever. In an effort to ensure reliability, distributed applications are generally built upon a reliable broadcast layer implemented by the Transmission Control Protocol (TCP) \cite{tanenbaum2011} -- one of the main protocols in the internet protocol suite. However, while TCP guarantees eventual delivery of all sent messages, it does not provide any upper time bound on delivery time \cite[p.~9]{freiling2011failure}. In practice, these assumptions do not hold simultaneously. 

In this work, we assume no such underlying structure exists and analyze the performance of a simple majority-rule protocol solving a fundamental coordination problem in distributed systems - \emph{binary majority consensus}, in the presence of probabilistic message loss. Using probabilistic analysis for a large scale, fully-connected, network of $2n$ agents, we prove that the Simple Majority Protocol (SMP) converges rapidly to a consensus on the majority opinion of the network with probability approaching $1$ as $n \rightarrow \infty$, given that the difference between the numbers of agents that hold different opinions grows as fast as $\sqrt{n}$. 
Otherwise, if the difference between the numbers of agents that hold different opinions is relatively close to zero, then the SMP still converges extremely fast to a consensus, but not necessarily on the initial majority opinion of the network.     

\subsection{Importance of Reliable Communication}

Reliability of communication is essential to guarantee coordination in almost all cases. The pitfalls and design challenges of coordination when communication is unreliable is best illustrated by \emph{the two generals' problem}, which was popularized by Jim Gray \cite{gray1978notes}:

Consider two generals who must coordinate a joint attack on an enemy. Both generals must attack simultaneously in order for the attack to succeed. While the two generals have agreed that they will attack, they haven't agreed upon a time for the attack. In order to coordinate, they can send messages to one another by running messengers. However, the messengers can be captured by the enemy and their messages will never reach their destination. 

Due to the uncertainty of message delivery, there exists no deterministic joint communication protocol which guarantees coordinated attack. To see this, assume there exists such a protocol by contradiction. Since a deterministic protocol must solve the problem in a finite number of steps, then the protocol prescribes a fixed number of message exchanges between the two generals, after which both must attack together. Some of these messages are successfully delivered and some are lost. Consider the last successfully delivered message in a run of the protocol, after which the recipient is confident enough to attack without the need for any further correspondence. Suppose this message was lost instead, then the recipient will hold off and not attack. However, the sender does not know about this last communication failure. By the protocol definition he must attack anyway, despite his counterpart's reluctance — contradicting the assumption that the protocol was a solution to the problem. 

\subsection{Majority Consensus}
The impossibility result of the two generals problem has had far-reaching implications in the field of distributed protocols and databases, including the study of binary consensus \cite{fischer1983consensus}. In the binary consensus problem, every agent is initially assigned some binary value, referred to as the agent's initial opinion. The goal of a protocol that solves consensus is to have every agent eventually decide on the same opinion, thus reaching agreement throughout the system. More formally, given any initial assignment of agent opinions, a run of a protocol which solves consensus must exhibit the following three properties:
\begin{enumerate}
	\item \textbf{Decision} Every agent eventually decides on some opinion $v \in \{0,1\}$.
	\item \textbf{Agreement} If some agent has decided on $v$, no opinion other than $v$ can be decided on by any other agent.
	\item \textbf{Non-Triviality} If some agent has decided on $v$, then $v$ was an opinion initially assigned to some agent.
\end{enumerate}

Consensus is a fundamental problem in distributed systems, as many other coordination problems have been shown to be directly reducible to and from consensus. The list includes agreeing on what transactions to commit to a database \cite{gray2006consensus}, state machine replication \cite{antoniadis2018state}, atomic snapshots \cite{attiya1998atomic}, total ordering of concurrent events \cite{lamport2019time}, and the two generals' problem, implying that no protocol can guarantee all three properties when communication is unreliable \cite{lynch1996distributed}. 

In light of this, it is interesting to consider a variation of the two generals' problem where the probability of a messenger getting captured is $p$  (independently of other messengers) \cite{halpern1993knowledge, rubinstein1989electronic}. While coordinated attack is still deterministically impossible, it is straightforward to design a protocol that guarantees success with probability at least $q$, which can be as close as desired to $1$. The first general simply sends $\lceil log_p(1-q) \rceil$ messengers, then attacks at the specified time without waiting for a reply, and the second general attacks if any messenger from the first general arrives.

In this work, we investigate whether leveraging such an assumption helps to solve binary majority consensus, in which the non-triviality clause stipulates that if a majority of agents initially hold the same opinion, then all agents must decide on this opinion. This variant of consensus is utilized when the agreed upon opinion holds importance beyond facilitating agreement. For example, a distributed system of sensors capable of detecting natural gas could use majority consensus to answer the question ``Is the amount of gas in the air greater than 10,000 ppm?'' in order to help detect a gas leak in a gas processing center. In this case, the opinion of a majority of the sensors would be the most trustworthy \cite{al2013binary}.

We analyze the performance of the SMP in a complete graph of communication, i.e., where each agent has an active communication channel to every other agent in the system. In SMP, agents communicate in equal-length time intervals called rounds. All messages are sent at the beginning of a communication round, and either arrive by the end of the round or are considered lost. We assume that all message loss events are i.i.d. with some constant probability. 

The SMP can be briefly described as follows: In each round, every agent sends its current opinion to all other agents. Then, it waits to receive all messages from other agents proposing their own opinions. If a majority of received messages propose the same opinion, then the agent adopts this opinion for the next round. All ties are reconciled by readopting the agent's own opinion. After a fixed number of rounds \textbf{\emph{r}}, each agent decides on its currently adopted opinion.

Similarly to the probabilistic protocol for the two generals' problem discussed above, the SMP does not solve consensus deterministically, but rather provides probabilistic guarantees instead. The \textbf{Decision} and \textbf{Non-Triviality} properties of classical consensus are assured, since all agents decide by the end of round \textbf{\emph{r}} and any opinion that was decided on, was proposed by some agent. However, \textbf{Agreement} is not assured, since there always exists a non-zero probability of a run of the protocol in which message losses cause one agent to see only one opinion and another agent to see only the other, thus making them disagree. Likewise, \textbf{Non-Triviality} of \emph{majority} consensus is not guaranteed, since the majority opinion could be hidden from some agent. We will show in this article that the probability of these runs is negligible as the number of agents, $n$, tends to infinity, and thus demonstrate that unreliable communication is not an insurmountable obstacle for coordination.

Specifically, we prove that the SMP with $\boldsymbol{r}=3$ reaches classical consensus with probability converging to 1 as $n$ tends to infinity. In a system of $2n$ agents, let $\delta_n$ be the the number of agents that are initially assigned the majority opinion\footnote{For simplicity, assume the majority opinion is always the same for all $n$.} minus $n$. We show that if $\delta_n$ grows at a rate of $\sqrt{n}$, then the SMP with $\boldsymbol{r}=2$ reaches majority consensus with probability approaching $1$ as $n \rightarrow \infty$.    
We also show that if $\delta_n$ grows at a rate faster than $\sqrt{n}$, then the SMP with $\boldsymbol{r}=1$ reaches majority consensus with probability that converges to 1 exponentially fast.

We also show that these achievability results are, in fact, tight. We will prove that if $\delta_n = 0$, then $\boldsymbol{r}=3$ communication rounds is a necessary condition, since the probability to reach consensus with only $\boldsymbol{r}=2$ rounds converges to 0 as $n \to \infty$.    
Similarly, if $\delta_n$ grows as slow as $\sqrt{n}$, then $\boldsymbol{r}=2$ communication rounds are a necessary condition to reach majority consensus.

\subsection{Related Work}
The problem of binary majority consensus has been extensively researched in many different fields and contexts including autonomous systems \cite{gacs1978one, mustafa2001majority, moreira2004efficient, gogolev2015distributed}, distributed systems \cite{thomas1979majority, breitwieser1982distributed, kanrar2016new} and information theory \cite{mostofi2007binary, perron2009using, cruise2014probabilistic}. 
Almost always the problem is studied in the context of possible failure of some aspect of the network. In distributed systems, failure most often arises from agents behaving maliciously, failing to follow the protocol, or outright crashing. Consequently, protocols that solve consensus (and majority consensus by extension) are designed to tolerate a certain fraction of the set of agents failing \cite{wensley1978sift,pease1980reaching}. Transmission faults (i.e., message loss, erasure or addition) can be considered an extension of agent failure, but doing so may lead to false conclusions. For example, in a system of $n$ agents, the entire system may be considered faulty even if only one message from each agent is lost. However, as shown by Santoro and Widmayer \cite{santoro1989time}, the system may tolerate up to $n-1$ messages losses in a round and still reach consensus. Additionally, assuming a probability distribution on message loss is consistent with how network protocols are analyzed. The most notable example is that TCP throughput has been shown to be inversely proportional to the square root of the link's average packet (i.e., message) loss probability \cite{padhye1998modeling}. 

In \cite{gacs1978one, moreira2004efficient, gogolev2015distributed, mostofi2007binary}, the authors studied the effects of message loss, random topology, Gaussian noise, and faulty agents,
on the SMP's convergence rate, i.e., the fraction of initial assignments of agent opinions (out of $2^n$) resulting in successful agreement. Specifically, in \cite{gogolev2015distributed} computer simulations showed an improvement in the convergence rate of the SMP as the message loss probability increased up to $0.8$, after which the rate begins to decrease to zero. In contrast, we are interested in the maximal probability of failure over any initial assignment of agent opinions, since we cannot assume any distribution or frequency on the input to the consensus problem.

Mustafa and Peke{\v{c}} \cite{mustafa2001majority}, studied the requirements on the connectivity of the network such that, under assumption of reliable communication, SMP achieves consensus on any initial assignment of agent opinions. Their main result is that the SMP computes the majority consensus successfully only in highly-connected networks. This conclusion led us to analyze the SMP under the assumption of a fully-connected network. However, message loss may actually improve the chances of consensus in graphs with lesser degrees of connectivity, as shown in \cite{gogolev2015distributed}. We leave the proof of this hypothesis to future work. Additionally, the  complete graph assumption is a valid approximation for unstructured overlays in peer to peer networks, e.g., Freenet, Gnutella and Fast Track \cite{lua2005survey}.

Our work closely resembles the work done in \cite{perron2009using, cruise2014probabilistic}. These articles have shown that in a lossless fully-connected network where agents poll a portion of their neighbors uniformly at random, the SMP converges quickly to majority consensus with probability of error (in the sense that agreement was reached, but not on the majority opinion) that decays exponentially with $n$. While assuming the existence of infinite agents in a system may initially seem ludicrous and impractical, our own computer simulations of the SMP have shown that these kind of results hold true even if the number of agents is on order of $10^6$, which is already the case in cryptocurrency protocols.  We add another assumption of unreliable communication and show that this, essentially, does not change the outcome.

The remaining part of the paper is organized as follows. 
In Section 2, we establish notation conventions. 
In Section 3, we formalize the model, the protocol, and the objectives of this work. 
In Section 4, we provide and discuss the main results of this work, and in Section 5, we prove them.

\section{Notation Conventions}

Throughout the paper, random variables will be denoted by capital letters, realizations will be denoted by the corresponding lower case letters, and their alphabets will be denoted by calligraphic letters. 
Random vectors and their realizations will be denoted, 
respectively, by boldface capital and lower case letters. 
Their alphabets will be superscripted by their dimensions. 
The binary Kullback--Leibler divergence function between two binary probability distributions with parameters $\alpha,\beta \in [0,1]$ is defined as
\begin{align} \label{DEF_Bin_DIVERGENCE}
D(\alpha \| \beta) = \alpha \log \left(\frac{\alpha}{\beta}\right) + (1-\alpha) \log \left(\frac{1-\alpha}{1-\beta}\right),
\end{align}
where logarithms, here and throughout the sequel, are understood to be taken to the natural base.
The cumulative distribution function of a standard normal random variable is defined by
\begin{align} \label{DEF_PHI}
\Phi(t) = \int_{-\infty}^{t} \frac{1}{\sqrt{2\pi}} \exp\left\{-\frac{s^{2}}{2}\right\} ds.
\end{align}
The probability of an event $\mathcal{E}$ will be denoted by $\pr \{ \mathcal{E} \}$, and the expectation operator w.r.t.\ a 
probability distribution $Q$ will be denoted by $\mathbb{E}_{Q} [\cdot]$, where the subscript will often be omitted.
The variance of a random variable $X$ is denoted by $\textbf{Var}[X]$. 
The indicator function of an event $\calA$ 
will be denoted by $\IND\{\calA\}$. 
The set $\{1,2,\ldots,n\}$ will often be denoted by $[1:n]$. 
For $\bx = (x_{1},x_{2},\ldots,x_{n}) \in \calX^{n}$ and for any $a \in \calX$, let us denote
\begin{align}
N(\bx;a) = \sum_{i=1}^{n} \IND\{x_{i}=a\}.
\end{align}

For two non-negative sequences ${a_{n}}$ and ${b_{n}}$, the sequence $A_{n}=n+a_{n}$ is called asymmetric of exact order of $b_{n}$ if there exists some $\alpha > 0$ such that $\lim_{n \to \infty} \frac{a_{n}}{b_{n}} = \alpha$. Also, the sequence $A_{n}=n+a_{n}$ is called asymmetric of order larger than $b_{n}$ if $\lim_{n \to \infty} \frac{a_{n}}{b_{n}} = \infty$.


\section{Model, Protocol, and Objectives}

Assume a set of $2n$ agents, and denote their assignment of initial opinions by $\bx_{0,n} \in \{0,1\}^{2n}$. The vector $\bx_{0,n}$ is called the initial state. Denote the numbers of zeros and ones in $\bx_{0,n}$ by $\mathsf{I}_{0}$ and $\mathsf{I}_{1}$, respectively. At each round, each agent transmits its current state to all other agents. 
If a message sent between any pair of agents arrives, then it is assumed to be delivered correctly. Otherwise, if $x\in \{0,1\}$ is transmitted between any pair of agents, but got lost, then the designated receiver receives the default symbol $e$\footnote{This assumption is only made for the purpose of making the definitions that follow brighter.}. For a sent message $x\in \{0,1\}$ and a received message $Y\in \{0,e,1\}$, we assume that all message losses are statistically independent and identically distributed according to\footnote{The binary erasure channel is characterized by a similar conditional distribution, but note that the actual faults in our model are message losses, not to be confused with erasures, which are different kinds of faults.} $\pr(Y=0|x=0)=\pr(Y=1|x=1)=1-q$ and $\pr(Y=e|x=0)=\pr(Y=e|x=1)=q$, where $q \in [0,1]$ is the loss parameter of the network. The two extreme cases of a reliable network (i.e., with $q=0$) and a completely unreliable network (i.e., with $q=1$) are of less interest, for obvious reasons, hence we assume throughout that $q \in (0,1)$.      


At round $\ell \geq 1$, the agent $i \in [1:2n]$ receives the (random) vector: 
\begin{align}
\by_{\ell}^{i} = (y_{\ell}^{i}(1),y_{\ell}^{i}(2), \ldots, y_{\ell}^{i}(i-1), y_{\ell}^{i}(i+1), \ldots, y_{\ell}^{i}(2n)) \in \{0,e,1\}^{2n-1},
\end{align}
and for $a \in \{0,1\}$, he calculates the enumerators:
\begin{align}
\mathsf{N}_{\ell,i}(a) = \IND\{\bx_{\ell-1}(i) = a\} + \sum_{j \neq i} \IND\{y_{\ell}^{i}(j) = a\}.
\end{align}
In the SMP, each agent updates its value according to the more common value at hand, i.e., agent $i$ chooses:
\begin{align}
\bx_{\ell}(i)
= \left\{   
\begin{array}{l l}
0    & \quad \text{if        $\mathsf{N}_{\ell,i}(0) > \mathsf{N}_{\ell,i}(1)$}   \\
1    & \quad \text{if        $\mathsf{N}_{\ell,i}(0) < \mathsf{N}_{\ell,i}(1)$}   \\
\bx_{\ell-1}(i)    & \quad \text{if        $\mathsf{N}_{\ell,i}(0) = \mathsf{N}_{\ell,i}(1)$}
\end{array} \right. .
\end{align} 
The vector $\bx_{\ell} \in \{0,1\}^{2n}$ is called the state at the end of round $\ell$.   

A specific SMP defines a-priori the number of rounds until termination. 
Let us denote by SMP$(r)$ the SMP with $r$ rounds of communication until termination.
We say that the SMP$(r)$ attains {\it consensus} if 
\begin{align}
\bx_{r}(1) = \bx_{r}(2) = \ldots = \bx_{r}(2n), 
\end{align}  
and denote this event by $\calC_{n}$. 
Similarly, we say that the SMP$(r)$ attains {\it majority consensus} if the following holds: 
\begin{align}
\mathsf{I}_{0} > \mathsf{I}_{1} ~~ &\rightarrow ~~ 
\bx_{r}(1) = \bx_{r}(2) = \ldots = \bx_{r}(2n) = 0, \\
\mathsf{I}_{0} < \mathsf{I}_{1} ~~ &\rightarrow ~~ 
\bx_{r}(1) = \bx_{r}(2) = \ldots = \bx_{r}(2n) = 1, \\
\mathsf{I}_{0} = \mathsf{I}_{1} ~~ &\rightarrow ~~ 
\bx_{r}(1) = \bx_{r}(2) = \ldots = \bx_{r}(2n),
\end{align}  
and denote this event by $\calC_{n}^{\mbox{\tiny m}}$.

For a specific initial state $\bx_{0,n}$, the probability of error in achieving consensus is defined as $P_{\mbox{\tiny e}}(\bx_{0,n}) = \pr[\calC_{n}^{\mbox{\tiny c}}]$. The maximal error probability with respect to the initial state is defined by
\begin{align}
P_{\mbox{\tiny e,max}} = \max_{\bx_{0,n} \in \{0,1\}^{2n}} P_{\mbox{\tiny e}}(\bx_{0,n}).
\end{align} 
The error probability in achieving majority consensus is defined similarly and denoted $P_{\mbox{\tiny e}}^{\mbox{\tiny m}}(\bx_{0,n})$. 

Now, the first objective of this work is to prove that the SMP requires only very few rounds of communication in order to attain consensus, with a maximal error probability that converges to 0 when $n \to \infty$. The second objective is to determine for which initial states it is possible to also achieve majority consensus with a small probability of error.

\section{Main Results}

The first main result of this work is the following, which is proved in Subsection \ref{SEC5.1}.

\begin{theorem} \label{Main_THEOREM}
	Let $\{\bx_{0,n}\}_{n \geq 1}$, be a sequence of initial states over $2n$ agents.
	Assume that the $2n$ agents communicate over a network with a loss parameter $q \in (0,1)$. Then,
	\begin{enumerate}
		\item If $\{\bx_{0,n}\}_{n \geq 1}$ is asymmetric of order larger than $\sqrt{n}$, the SMP$(1)$ attains $\pr[\calC_{n}^{\mbox{\tiny m}}] \xrightarrow{n \to \infty} 1$.
		\item If $\{\bx_{0,n}\}_{n \geq 1}$ is asymmetric of exact order of $\sqrt{n}$, the SMP$(2)$ attains $\pr[\calC_{n}^{\mbox{\tiny m}}] \xrightarrow{n \to \infty} 1$.
		\item For any $\{\bx_{0,n}\}_{n \geq 1}$, the SMP$(3)$ attains $\pr[\calC_{n}] \xrightarrow{n \to \infty} 1$.
	\end{enumerate}
\end{theorem}
We now provide a short discussion on the results of Theorem \ref{Main_THEOREM}. 

As can be seen in Theorem \ref{Main_THEOREM}, the SMP requires at most three rounds of communications in order to attain consensus, in the limit of an infinite number of agents. Consensus on the majority cannot be ensured for all possible initial states, but only for those initial states that have a significant majority to one of the sides. In order to understand this fact better, consider the following special case. Assume a network with $2n$ agents, such that $\mathsf{I}_{0} = n + \log(n)$ and $\mathsf{I}_{1} = n - \log(n)$. Since this majority in favor of the zeros is so weak, then it is most likely that the random losses in the network will completely hide it; we expect that about half of the agents will have $\mathsf{N}_{1,i}(0) > \mathsf{N}_{1,i}(1)$, thus updating their current opinion to `$0$', while the other half will update their current opinion to `$1$'s. We conclude that the state at the end of round 1 is probabilistically equivalent to a sequence of $2n$ fair coin tosses, and hence, with a probability of about one half, the majority at the end of round 1 will be different from the initial majority.               

More quantitatively, let $\mathsf{I}_{0} = n + a_{n}$ and $\mathsf{I}_{1} = n - a_{n}$, where $\{a_{n}\}_{n \geq 1}$ is a non-negative non-decreasing sequence. Also, for an agent with an initial opinion `0', let $p_{n}$ denote the sequence of probabilities of the events that such an agent updates its opinion to `0'. Then, the following trichotomy is seen inside the proof of Theorem \ref{Main_THEOREM}:
\begin{fact} \label{Trichotomy}
	The following trichotomy holds.
\begin{enumerate}
	\item If $\lim_{n \to \infty} \frac{a_{n}}{\sqrt{n}} = 0$, then $p_{n} \xrightarrow{n \to \infty} \frac{1}{2}$.  
	\item If $\lim_{n \to \infty} \frac{a_{n}}{\sqrt{n}} = \alpha \in (0,\infty)$ then $p_{n} \xrightarrow{n \to \infty} \beta(\alpha,q) \in (\tfrac{1}{2},1)$.
	\item If $\lim_{n \to \infty} \frac{a_{n}}{\sqrt{n}} = \infty$, then $p_{n} \xrightarrow{n \to \infty} 1$.
\end{enumerate}  	
\end{fact}
One of the most surprising facts, at least to the authors of this work, is the following. For highly symmetric initial states, although $p_{n} \xrightarrow{n \to \infty} \frac{1}{2}$ (which is proved in Appendix C), it turns out (see Proposition \ref{PROP_3} in Subsection \ref{SEC5.1}) that after a single round of communication, the initial symmetry breaks equiprobably into one of the sides. 
Moreover, for the symmetric case of $\mathsf{I}_{0} = \mathsf{I}_{1}=n$, we prove in Propositions \ref{PROP_3} and \ref{PROP_4} that with a probability converging to 1, the state at the end of round 1 will be asymmetric of exact order of $\sqrt{n}$. Then, according to the second point in Fact \ref{Trichotomy}, the state at the end of round 2 is going to have a significant majority to one of the sides, and thus, according to the third point in Fact \ref{Trichotomy}, only one more round of communication is required in order to achieve consensus. One should note that if the initial state is already asymmetric of exact order of $\sqrt{n}$, then only two rounds of communication are needed for attaining consensus, and in this case, it is guaranteed (with high probability) that all agents agree on the initial majority opinion.        

The phenomenon that the initial symmetry breaks into a sufficient majority after the first round is of key importance, since it makes the convergence of the SMP so rapid. In fact, we also conclude that the faulty communication between the agents even helps in attaining consensus, by breaking the symmetry in some extreme cases. E.g., consider the case of $\mathsf{I}_{0}=\mathsf{I}_{1}=n$ and a reliable network (i.e., the case of $q=0$). Then, ad infinitum, the state at the end of any round will be symmetric. Otherwise when losses exist according to some $q \in (0,1)$, this will not be the case, even if the percentage of losses is extremely small (but fixed at all $n$).        

A significant difference exists between the first point of Theorem \ref{Main_THEOREM} and its last two points, which is the following. The first point of Theorem \ref{Main_THEOREM} is based on Proposition 1 in Subsection \ref{SEC5.1}, which is mainly proved by using the Chernoff bound. Since the Chernoff bound is a non-asymptotic tool, we acquire a large-deviations result, i.e., for a given sequence $\{a_{n}\}_{n \geq 1}$ (with the condition $\lim_{n \to \infty} \frac{a_{n}}{\sqrt{n}} = \infty$), we propose a tight\footnote{This tightness follows from the fact that a lower bound with a matching exponent can be derived as well.} upper bound on $P_{\mbox{\tiny e}}^{\mbox{\tiny m}}(\bx_{0,n})$, which holds for any finite $n$. This result is obviously stronger than just $\pr[\calC_{n}^{\mbox{\tiny m}}] \xrightarrow{n \to \infty} 1$.  
On the other hand, the second and the third points of Theorem \ref{Main_THEOREM} are based on Propositions 2 and 3 in Subsection \ref{SEC5.1}, respectively. Since the proofs of these propositions involve central limit theorems, we merely arrive at asymptotic results. As a consequence, we do not know at what rates the probabilities in the second and the third points of Theorem \ref{Main_THEOREM} converge to one.            

Since the results of the second and the third points of Theorem \ref{Main_THEOREM} are merely asymptotic, a few words on finite $n$ effects are in order. We base the following facts on computer simulations of the SMP. On the one hand, convergence to consensus at more than three rounds is definitely possible, but only when the initial state is symmetric or almost symmetric. The reason for that is the fact mentioned above, according to which, the state at round 1 is probabilistically equivalent to a sequence of $2n$ fair coin tosses, and hence, the probability that the state at round 1 is again symmetric behaves asymptotically\footnote{Upper and lower bounds can be derived using the Stirling's bounds to $n!$.} as $1/\sqrt{n}$, which is not negligible at all, even for a relatively large number of agents. 
For relatively small values of $n$, we observed several realizations with even more than a single returning to a fully symmetric state. Although quite rare, these events should be taken into consideration in practical implementations.    


All the results provided in Theorem \ref{Main_THEOREM} are, in fact, achievability results, i.e., they only tell under what conditions consensus can be attained. Hence, it is worth investigating whether consensus may be attained by the SMP with even less communication rounds than required in Theorem \ref{Main_THEOREM}.     
In the following result, which is the second main result of this work and is proved in Subsection \ref{SEC5.2}, we show that for highly symmetric initial states, three rounds of communications are not only sufficient, but also necessary.   

\begin{theorem} \label{THEOREM2}
	Let $\{\bx_{0,n}\}_{n \geq 1}$, be a sequence of symmetric initial states over $2n$ agents, i.e., $N(\bx_{0,n};0)=N(\bx_{0,n};1)=n$ for all $n$. 
	Assume that the $2n$ agents communicate over a network with a loss parameter $q \in (0,1)$. Then, the SMP$(2)$ attains $\pr[\calC_{n}] \xrightarrow{n \to \infty} 0$.
\end{theorem}

While Theorem \ref{THEOREM2} provides a converse result with regard to the third point of Theorem \ref{Main_THEOREM}, a similar converse result can also be established with regard to the second point of Theorem \ref{Main_THEOREM}. If the initial state is asymmetric of exact order of $\sqrt{n}$, then the SMP will likely not attain consensus after only a single round of communication, and furthermore, the probability of reaching consensus will tend to 0 as $n \to \infty$. We omit the proof of this negative result.

\section{Proofs} 
\label{SEC5}
\subsection{Proof of Theorem \ref{Main_THEOREM}}
\label{SEC5.1}
The first point of Theorem \ref{Main_THEOREM} is proved via the following result, which is proved in Appendix A.

\begin{proposition} \label{PROP_1}
	Let $\{A_{n}\}_{n=1}^{\infty}$ be a sequence such that $\lim_{n \to \infty} \frac{A_{n}}{\sqrt{n}} = \infty$. For an initial state $\bx_{0,n} \in \{0,1\}^{2n}$ with at least $n+A_{n}$ zeros or at least $n+A_{n}$ ones and a channel parameter $q \in [0,1)$, the SMP$(1)$ attains $\pr[\calC_{n}^{\mbox{\tiny m}}] \xrightarrow{n \to \infty} 1$. Specifically, if $\lim_{n \to \infty} \frac{A_{n}}{n} < 1$, then 
	\begin{align}
	P_{\mbox{\tiny e}}^{\mbox{\tiny m}}(\bx_{0,n})
	&\leq 2n \sqrt{\frac{n+A_{n}}{n-A_{n}}} \cdot \exp \left\{-(1-q) \cdot \frac{A_{n}^{2}}{n} \right\}.
	\end{align}
\end{proposition}

In order to prove the second point of Theorem \ref{Main_THEOREM}, we rely on the following result, which is proved in Appendix B. 

\begin{proposition} \label{PROP_2}
	Let $q \in [0,1)$ be a channel parameter. 
	Let $\alpha > 0$ be fixed and let $0 < \epsilon < \Phi(t_{0})-\tfrac{1}{2}$, where $t_{0} = \sqrt{2\alpha^{2}(1-q)/q}$.	 
	Then, the SMP$(1)$ attains the following. 
	\begin{enumerate}
		\item If $\bx_{0,n} \in \{0,1\}^{2n}$ has at least $n+\alpha \sqrt{n}$ zeros, then
		\begin{align}
		\pr \left\{N(\bX_{1};0) \geq 2n (\Phi(t_{0})-\epsilon) \right\}
		\xrightarrow{n \to \infty} 1.
		\end{align}
		\item If $\bx_{0,n} \in \{0,1\}^{2n}$ has at least $n+\alpha \sqrt{n}$ ones, then
		\begin{align}
		\pr \left\{N(\bX_{1};1) \geq 2n (\Phi(t_{0})-\epsilon) \right\}
		\xrightarrow{n \to \infty} 1.
		\end{align} 
	\end{enumerate}
\end{proposition}
Then, combining the results of Propositions \ref{PROP_1} and \ref{PROP_2} using the law of total probability, the second point of Theorem \ref{Main_THEOREM} follows immediately.
 
In order to prove the third point of Theorem \ref{Main_THEOREM}, we provide one more result.
The following proposition shows that if the initial state is symmetric, then the state at round one will be asymmetric of order at least $\sqrt{n}$. This result is proved in Appendix C.

\begin{proposition} \label{PROP_3}
	Let $\bx_{0,n} \in \{0,1\}^{2n}$ be an initial state with $n$ zeros and $n$ ones and let $q \in (0,1)$ be a channel parameter. Let $\epsilon > 0$ be given. Then, there exist $\delta=\delta(\epsilon)$ with $\delta(\epsilon) \xrightarrow{\epsilon \to 0} 0$ and $M(\epsilon)$, such that for all $n \geq M(\epsilon)$,
	\begin{align}
	\pr \left\{ \{N(\bX_{1};0) \leq n - \delta\sqrt{n}\}
	\cup \{N(\bX_{1};0) \geq n + \delta\sqrt{n}\} \right\} \geq 1 - \epsilon.
	\end{align} 
\end{proposition}

We are now able to prove the third point of Theorem \ref{Main_THEOREM}. 
Let $\epsilon_{1},\epsilon_{3} > 0$ be given, and let $\delta$ be as in Proposition \ref{PROP_3} corresponding to $\epsilon_{3}$.
Also, let $t_{0} = \sqrt{2\delta^{2}(1-q)/q}$, choose $\epsilon_{2}>0$ such that $\Phi(t_{0})-\epsilon_{2} > 1/2$, and denote $\beta=2(\Phi(t_{0})-\epsilon_{2})-1$.  
Define the following events
\begin{align}
\calA_{n} = \left\{N(\bX_{1};0) \leq n - \delta \sqrt{n} \text{~~or~~} N(\bX_{1};0) \geq n + \delta \sqrt{n} \right\},
\end{align}
and
\begin{align}
\calB_{n} = \left\{N(\bX_{2};0) \leq (1-\beta)n \text{~~or~~} N(\bX_{2};0) \geq (1+\beta)n \right\}.
\end{align} 
Then, consider the following.
\begin{align}
\pr\{\calC_{n}\} 
&= \pr\{ N(\bX_{3};0) = 0 \text{~or~} N(\bX_{3};0) = 2n\}\\
\label{A_To_exp1}
&= \pr\{ N(\bX_{3};0) = 0 \text{~or~} N(\bX_{3};0) = 2n| \calB_{n}\} \cdot \pr\{\calB_{n}\} \nn \\
&~~+ \pr\{ N(\bX_{3};0) = 0 \text{~or~} N(\bX_{3};0) = 2n| \calB_{n}^{\mbox{\tiny c}}\} \cdot \pr\{\calB_{n}^{\mbox{\tiny c}}\}  \\
&\geq \pr\{ N(\bX_{3};0) = 0 \text{~or~} N(\bX_{3};0) = 2n| \calB_{n}\} \cdot \pr\{\calB_{n}\}  \\
\label{A_To_exp2}
&\geq \left(1 - \epsilon_{1}\right) \cdot \pr\{\calB_{n}\},
\end{align}
where \eqref{A_To_exp1} follows from the law of total probability and \eqref{A_To_exp2} holds for all large enough $n$, due to Proposition \ref{PROP_1}. Furthermore,
\begin{align}
\pr\{\calB_{n}\} 
\label{A_To_exp3}
&= \pr\{ \calB_{n}| \calA_{n}\} \cdot \pr\{\calA_{n}\} + \pr\{ \calB_{n}| \calA_{n}^{\mbox{\tiny c}}\} \cdot \pr\{\calA_{n}^{\mbox{\tiny c}}\}  \\
&\geq \pr\{ \calB_{n}| \calA_{n}\} \cdot \pr\{\calA_{n}\} \\
\label{A_To_exp4}
&\geq (1-\epsilon_{2}) \cdot \pr\{\calA_{n}\} \\
\label{A_To_exp5}
&\geq (1-\epsilon_{2}) \cdot (1-\epsilon_{3}),
\end{align}
where \eqref{A_To_exp3} is again due to the law of total probability, \eqref{A_To_exp4} follows from Proposition \ref{PROP_2} for all $n$ sufficiently large, and \eqref{A_To_exp5} follows from Proposition \ref{PROP_3}, also for all $n$ sufficiently large. Substituting \eqref{A_To_exp5} back into \eqref{A_To_exp2}, we conclude that $\pr\{\calC_{n}\}$ can be made arbitrarily close to 1, which implies the result in the third point in Theorem \ref{Main_THEOREM}.

\subsection{Proof of Theorem \ref{THEOREM2}}
\label{SEC5.2}

The following proposition, which is proved in Appendix D, shows that if the initial state is symmetric, then the state at round one cannot be asymmetric of order larger than $\sqrt{n}$.  

\begin{proposition} \label{PROP_4}
	Let $\{B_{n}\}_{n=1}^{\infty}$ be a sequence such that $\lim_{n \to \infty} \frac{B_{n}}{\sqrt{n}} = \infty$. For an initial state $\bx_{0,n} \in \{0,1\}^{2n}$ with $n$ zeros and $n$ ones and a channel parameter $q \in (0,1)$, the following holds:
	\begin{align}
	\pr \left\{ \{N(\bX_{1};0) \leq n - B_{n}\}
	\cup \{N(\bX_{1};0) \geq n + B_{n}\} \right\} \leq 2\exp \left\{-\frac{B_{n}^{2}}{n} \right\}.
	\end{align} 
\end{proposition}
We also have the following result, which is proved in Appendix E.

\begin{proposition} \label{PROP_5}
	Let $\{C_{n}\}_{n=1}^{\infty}$ be a sequence such that $\lim_{n \to \infty} \frac{C_{n}}{n} = 0$. Let $\bx_{0,n} \in \{0,1\}^{2n}$ be an initial state with $n+C_{n}$ zeros or $n+C_{n}$ ones. Let $q \in (0,1)$ be a channel parameter and denote the constant $f_{q}=32/\min\{q,1-q\}$. 
	Then, the SMP$(1)$ is characterized by
	\begin{align}
	\pr\{\calC_{n}\} 
	\leq \exp\left\{- C_{n}^{2} \cdot
	\exp \left\{- f_{q} \cdot  \frac{C_{n}^{2}}{n-C_{n}} \right\} \right\}.
	\end{align}
\end{proposition}
We are now in a good position to prove Theorem 2. 
Let $C(q)=\frac{1}{2} f_{q}^{-1}$, choose the sequence
\begin{align}
\Theta_{n}=\sqrt{C(q) n \log(n)},
\end{align} 
and define the sequence of events
\begin{align}
\calF_{n} = \{N(\bX_{1};0) \leq n - \Theta_{n}\}
\cup \{N(\bX_{1};0) \geq n + \Theta_{n}\}.
\end{align}
According to Proposition \ref{PROP_4}, we have that
\begin{align}
\pr \left\{ \calF_{n} \right\} 
&\leq 2\exp \left\{-\frac{C(q) n \log(n)}{n} \right\} \\
&= 2\exp \left\{- C(q) \log(n) \right\}\\
\label{Since1}
&= \frac{2}{n^{C(q)}},
\end{align}
which converges to zero as $n \to \infty$. 
In addition, it follows from Proposition \ref{PROP_5} that
\begin{align}
\pr\{ \calC_{n} | \calF_{n}^{\mbox{\tiny c}}\}
&\leq \exp\left\{- C(q) n \log(n)
\cdot \exp \left\{- f_{q} \cdot  \frac{C(q) n \log(n)}{n-\sqrt{C(q) n \log(n)}} \right\} \right\} \\
\label{ToExp4}
&\leq \exp\left\{- C(q) n \log(n)
\cdot \exp \left\{-  \frac{ \frac{1}{2} n \log(n)}{n-\frac{1}{2}n} \right\} \right\} \\
&= \exp\left\{- C(q) n \log(n) \cdot \exp \left\{- \log(n) \right\} \right\} \\
&= \exp\left\{- C(q) n \log(n) n^{-1} \right\} \\
&= \exp\left\{- C(q) \log(n) \right\} \\
\label{ToExp5}
&=  \frac{1}{n^{C(q)}},
\end{align}
where \eqref{ToExp4} holds for all large enough $n$.
Then, consider the following.
\begin{align}
\pr\{\calC_{n}\} 
\label{A_To_exp6}
&= \pr\{ \calC_{n} | \calF_{n}\} \cdot \pr\{\calF_{n}\} + \pr\{ \calC_{n} | \calF_{n}^{\mbox{\tiny c}}\} \cdot \pr\{\calF_{n}^{\mbox{\tiny c}}\}  \\
&\leq \pr\{\calF_{n}\} + \pr\{ \calC_{n} | \calF_{n}^{\mbox{\tiny c}}\} \\
\label{A_To_exp7}
&\leq \frac{2}{n^{C(q)}} + \frac{1}{n^{C(q)}}\\
&= \frac{3}{n^{C(q)}} \xrightarrow{n \to \infty} 0,
\end{align}
where \eqref{A_To_exp6} is due to the law of total probability and \eqref{A_To_exp7} follows from \eqref{Since1} and \eqref{ToExp5}.

\section*{Appendix A - Proof of Proposition \ref{PROP_1}}
\renewcommand{\theequation}{A.\arabic{equation}}
\setcounter{equation}{0}  

Due to symmetry, we only analyze the case $\mathsf{I}_{0} > \mathsf{I}_{1}$. 
It follows from the union bound that 
\begin{align}
P_{\mbox{\tiny e}}^{\mbox{\tiny m}}(\bx_{0,n})
&= \pr \left\{\bigcup_{i=1}^{2n} \{\bX_{1}(i) = 1\} \right\} \\
\label{ToSubs1}
&\leq \sum_{i=1}^{2n} \pr \left\{ \bX_{1}(i) = 1 \right\}.
\end{align} 

In the following, let us denote by $\text{Ber}(p)$ a Bernoulli random variable with a success probability $p$ and by $\text{Bin}(n,p)$ a binomial random variable with $n$ independent experiments, each one with a success probability $p$. We adopt the following convention: if an event contains at least 2 binomial random variables, then we assume that they are statistically independent.

Let us denote $q' = 1-q$.
If an agent starts with a `0', then the probability to decide in favor of `1' is upper-bounded by 
\begin{align}
\label{B_ToExp0}
&\pr\left\{\text{Bin}\left(n-A_{n},q'\right) \geq \text{Bin}\left(n+A_{n}-1,q'\right) +1+1 \right\}  \\
\label{B_ToExp1}
&~~\leq \pr\left\{\text{Bin}\left(n-A_{n},q'\right) \geq \text{Bin}\left(n+A_{n}-1,q'\right) + \text{Bin}\left(1,q'\right) \right\} \\
\label{ToCallA0}
&~~= \pr\left\{\text{Bin}\left(n-A_{n},q'\right) \geq \text{Bin}\left(n+A_{n},q'\right) \right\},
\end{align}
where the addition of the second 1 in \eqref{B_ToExp0} follows from the need to strictly break the tie in order to adopt `1' and \eqref{B_ToExp1} is due to the fact that $\text{Bin}\left(1,q'\right) \leq 2$ with probability one. 

If an agent starts with a `1', then the probability to decide `1' is upper-bounded by 
\begin{align}
&\pr\left\{\text{Bin}\left(n-A_{n}-1,q'\right) + 1 \geq \text{Bin}\left(n+A_{n},q'\right) \right\} \nn \\
\label{ToCallA1}
&~~\leq \pr\left\{\text{Bin}\left(n-A_{n},q'\right) + 1 \geq \text{Bin}\left(n+A_{n},q'\right) \right\}.
\end{align}
Since \eqref{ToCallA1} cannot be smaller than \eqref{ToCallA0}, we continue with \eqref{ToCallA1}.
From now on, we prove that the probability in \eqref{ToCallA1}, to be denoted by $P_{n}$, converges to zero as $n \to \infty$.
Let
\begin{align}
X_{n} = \sum_{\ell=1}^{n-A_{n}} I_{\ell},~~~ 
Y_{n} = \sum_{k=1}^{n+A_{n}} J_{k},
\end{align}
where $I_{\ell} \sim \text{Ber}(q')$, for all $\ell \in \{1,2,\ldots, n-A_{n}\}$, $J_{k} \sim \text{Ber}(q')$, for all $k \in \{1,2,\ldots, n+A_{n}\}$, and all of these binary random variables are independent.
Now,
\begin{align}
P_{n} 
&= \pr \{X_{n}+1 \geq Y_{n}\} \\
&= \pr \left\{e^{\lambda(X_{n}-Y_{n}+1)} \geq 1 \right\} \\
\label{ToCallA2}
&\leq \mathbb{E} \left[e^{\lambda(X_{n}-Y_{n}+1)}\right],  
\end{align}
where \eqref{ToCallA2} is due to Markov's inequality. Since \eqref{ToCallA2} holds for every $\lambda \geq 0$, it follows that
\begin{align}
P_{n}
\leq \inf_{\lambda \geq 0} \mathbb{E} \left[e^{\lambda(X_{n}-Y_{n}+1)}\right].
\end{align}
We get that
\begin{align}
\mathbb{E} \left[e^{\lambda(X_{n}-Y_{n}+1)}\right]
&= e^{\lambda} \cdot \mathbb{E}\left[\exp\left\{\lambda \left(\sum_{\ell=1}^{n-A_{n}} I_{\ell} - \sum_{k=1}^{n+A_{n}} J_{k}\right) \right\}\right] \\
&= e^{\lambda} \cdot \mathbb{E}\left[\prod_{\ell=1}^{n-A_{n}} e^{\lambda I_{\ell}} \cdot \prod_{k=1}^{n+A_{n}} e^{-\lambda J_{k}} \right] \\
\label{ToCallA3}
&= e^{\lambda} \cdot \prod_{\ell=1}^{n-A_{n}} \mathbb{E}\left[ e^{\lambda I_{\ell}} \right] \cdot \prod_{k=1}^{n+A_{n}} \mathbb{E}\left[ e^{-\lambda J_{k}} \right] \\
&= e^{\lambda} \cdot  \left[1+q'(e^{\lambda}-1)\right]^{n-A_{n}} \cdot \left[1+q'(e^{-\lambda}-1)\right]^{n+A_{n}}  \\
\label{ToCallA4}
&\leq e^{\lambda} \cdot  \left[\exp\{q'(e^{\lambda}-1) \}\right]^{n-A_{n}} \cdot \left[\exp\{q'(e^{-\lambda}-1)\}\right]^{n+A_{n}}  \\
\label{ToRef19}
&= \exp \left\{\lambda +q'(e^{\lambda}-1)(n-A_{n}) +q'(e^{-\lambda}-1)(n+A_{n}) \right\},
\end{align}
where \eqref{ToCallA3} is due to the independence of all binary random variables and \eqref{ToCallA4} follows from the inequality $1+x \leq e^{x}$. Upon defining 
\begin{align}
f(\lambda) = \lambda +q'(e^{\lambda}-1)(n-A_{n}) +q'(e^{-\lambda}-1)(n+A_{n}),
\end{align}
we find that 
\begin{align}
f'(\lambda) = 1 +q'e^{\lambda}(n-A_{n})  - q'e^{-\lambda}(n+A_{n}).
\end{align}
In order to facilitate expressions, we solve for $f'(\lambda)=1$ and find that  
\begin{align}
e^{\lambda^{*}} = \sqrt{\frac{n+A_{n}}{n-A_{n}}}.
\end{align}
Substituting it back into \eqref{ToRef19} yields that
\begin{align}
P_{n} 
&\leq \exp \left\{\lambda^{*} +q'(e^{\lambda^{*}}-1)(n-A_{n}) +q'(e^{-\lambda^{*}}-1)(n+A_{n}) \right\} \\
&= \sqrt{\frac{n+A_{n}}{n-A_{n}}} \cdot \exp \left\{q'\left(\sqrt{\frac{n+A_{n}}{n-A_{n}}}-1\right)(n-A_{n}) +q'\left(\sqrt{\frac{n-A_{n}}{n+A_{n}}}-1\right)(n+A_{n}) \right\} \\
&= \sqrt{\frac{n+A_{n}}{n-A_{n}}} \cdot \exp \left\{q'\left[\sqrt{(n+A_{n})(n-A_{n})}-n+A_{n}\right]  \right\} \nn \\
&~~~~~~~~~~~~~~~~~~~~~~~~~~~\times \exp \left\{q'\left[\sqrt{(n-A_{n})(n+A_{n})}-n-A_{n}\right]  \right\} \\
&= \sqrt{\frac{n+A_{n}}{n-A_{n}}} \cdot \exp \left\{2q'\left(\sqrt{n^{2}-A_{n}^{2}}-n\right) \right\}.
\end{align}
Consider the following
\begin{align}
\sqrt{n^{2}-A_{n}^{2}}-n
&= \sqrt{n^{2} \left(1-\frac{A_{n}^{2}}{n^{2}}\right)}-n \\
&= n \sqrt{1-\frac{A_{n}^{2}}{n^{2}}}-n \\
\label{ToCallA5}
&\leq n \left(1-\frac{A_{n}^{2}}{2n^{2}}\right)-n \\
&= -\frac{A_{n}^{2}}{2n},
\end{align}
where \eqref{ToCallA5} follows from the inequality $\sqrt{1-t} \leq 1-t/2$. Continuing from \eqref{ToSubs1}, we arrive at
\begin{align}
P_{\mbox{\tiny e}}^{\mbox{\tiny m}}(\bx_{0,n})
&\leq 2n \sqrt{\frac{n+A_{n}}{n-A_{n}}} \cdot \exp \left\{-(1-q) \cdot \frac{A_{n}^{2}}{n} \right\},
\end{align}
which converges to zero when $n \to \infty$, as long as $\lim_{n \to \infty} \frac{A_{n}}{\sqrt{n}} = \infty$ and $\lim_{n \to \infty} \frac{A_{n}}{n} < 1$.

For the case of $\lim_{n \to \infty} \frac{A_{n}}{n} = 1$, consider the following. Let $\{A_{n}\}_{n=1}^{\infty}$ be any sequence with $\lim_{n \to \infty} \frac{A_{n}}{n} = 1$ and let $\{A'_{n}\}_{n=1}^{\infty}$ be a sequence with $\lim_{n \to \infty} \frac{A'_{n}}{n} = \alpha$, for $\alpha \in (0,1)$. Then, for sufficiently large $n$, $A_{n} \geq A'_{n}$, and thus, it follows that 
\begin{align}
&\pr\left\{\text{Bin}\left(n-A_{n},q'\right) + 1 \geq \text{Bin}\left(n+A_{n},q'\right) \right\} \nn \\
&~~~~~~\leq \pr\left\{\text{Bin}\left(n-A'_{n},q'\right) + 1 \geq \text{Bin}\left(n+A'_{n},q'\right) \right\} \xrightarrow{n \to \infty} 0,
\end{align}
which completes the proof of Proposition \ref{PROP_1}.

\section*{Appendix B - Proof of Proposition \ref{PROP_2}}
\renewcommand{\theequation}{B.\arabic{equation}}
\setcounter{equation}{0}

\subsubsection*{Step 1: The Limit of the Probability to Decide `1'}


If an agent starts with a `1', then the probability to decide in favor of `1' is given by 
\begin{align}
\label{ToCall0}
&\pr\left\{\text{Bin}\left(n-\alpha\sqrt{n}-1,q'\right) + 1 \geq \text{Bin}\left(n+\alpha\sqrt{n},q'\right) \right\},
\end{align}
and if an agent starts with a `0', then the probability to decide in favor of `1' is given by 
\begin{align}
&\pr\left\{\text{Bin}\left(n-\alpha\sqrt{n},q'\right) \geq \text{Bin}\left(n+\alpha\sqrt{n}-1,q'\right) +2 \right\} \nn \\
\label{ToCall1}
&= \pr\left\{\text{Bin}\left(n-\alpha\sqrt{n},q'\right) \geq \text{Bin}\left(n,q'\right) + \text{Bin}\left(\alpha\sqrt{n}-1,q'\right) +2 \right\}.
\end{align}
From now on, we prove that the probability in \eqref{ToCall1}, to be denoted by $P_{n}$, converges to a value, which is strictly smaller than $\tfrac{1}{2}$ for all sufficiently large $n$.
An identical result also holds for the probability in \eqref{ToCall0}, the proof of which is very similar and hence omitted.
  
Let $I_{\ell} \sim \text{Ber}(q')$, for all $\ell \in \{1,2,\ldots, n-\alpha\sqrt{n}\}$, $J_{s} \sim \text{Ber}(q')$, for all $s \in \{1,2,\ldots, n\}$, as well as $K_{m} \sim \text{Ber}(q')$, for all $m \in \{1,2,\ldots, \alpha\sqrt{n}-1\}$, and all of these binary random variables are independent.
Consider the following
\begin{align}
P_{n} 
&= \pr\left\{\sum_{\ell=1}^{n-\alpha\sqrt{n}} I_{\ell} \geq \sum_{s=1}^{n} J_{s} + \sum_{m=1}^{\alpha\sqrt{n}-1} K_{m} +2 \right\}\\
&= \pr\left\{\sum_{\ell=1}^{n-\alpha\sqrt{n}} I_{\ell} - (n-\alpha\sqrt{n}+\alpha\sqrt{n})q' \geq \sum_{s=1}^{n} J_{s} -nq' + \sum_{m=1}^{\alpha\sqrt{n}-1} K_{m} +2 \right\}\\
&= \pr\left\{\sum_{\ell=1}^{n-\alpha\sqrt{n}} (I_{\ell}-q') - \alpha\sqrt{n}q' \geq \sum_{s=1}^{n} (J_{s} - q') + \sum_{m=1}^{\alpha\sqrt{n}-1} K_{m} +2 \right\}\\
&= \pr\left\{\frac{1}{\sqrt{n}}\sum_{\ell=1}^{n-\alpha\sqrt{n}} (I_{\ell}-q') \geq \frac{1}{\sqrt{n}}\sum_{s=1}^{n} (J_{s}-q') + \frac{1}{\sqrt{n}} \left(\sum_{m=1}^{\alpha \sqrt{n}-1} K_{m} +2 \right) + \alpha q' \right\}.
\end{align}
Let us denote 
\begin{align}
X_{n} = \frac{1}{\sqrt{n}}\sum_{\ell=1}^{n-\alpha\sqrt{n}} (I_{\ell}-q'),~~ Y_{n} = \frac{1}{\sqrt{n}}\sum_{s=1}^{n} (J_{s}-q'),~~ Z_{n} = \frac{1}{\sqrt{n}} \left(\sum_{m=1}^{\alpha \sqrt{n}-1} K_{m} + 2 \right).
\end{align}
It follows directly from the central limit theorem \cite[p.\ 112, Theorem 2.4.1.]{DURRETT} that $Y_{n}$ converges in distribution to $Y \sim \calN(0,\sigma^{2})$, where $\sigma^{2} = q'(1-q')$. 
Concerning the sequence $X_{n}$, we first write it as follows
\begin{align}
X_{n} 
&= \frac{1}{\sqrt{n}}\sum_{\ell=1}^{n-\alpha\sqrt{n}} (I_{\ell}-q') \\
&= \frac{\sqrt{n-\alpha\sqrt{n}}}{\sqrt{n}} \cdot \frac{1}{\sqrt{n-\alpha\sqrt{n}}}\sum_{\ell=1}^{n-\alpha\sqrt{n}} (I_{\ell}-q') \\
&\dfn \frac{\sqrt{n-\alpha\sqrt{n}}}{\sqrt{n}} \tilde{X}_{n}, 
\end{align}
where $\tilde{X}_{n}$ converges in distribution to $X \sim \calN(0,\sigma^{2})$, again, from the central limit theorem. 
In order to conclude that $X_{n}$ itself converges in distribution to $X \sim \calN(0,\sigma^{2})$, we only need to prove that $|\tilde{X}_{n}-X_{n}|$ converges in distribution to 0. We have that
\begin{align}
\lim_{n \to \infty} \mathbb{E} \left[\tilde{X}_{n}-X_{n}\right]^{2}
&= \lim_{n \to \infty} \mathbb{E}
\left[\left(1-\frac{\sqrt{n-\alpha\sqrt{n}}}{\sqrt{n}}\right) \cdot \frac{1}{\sqrt{n-\alpha\sqrt{n}}}\sum_{\ell=1}^{n-\alpha\sqrt{n}} (I_{\ell}-q') \right]^{2} \\
&= \lim_{n \to \infty} \left(1-\frac{\sqrt{n-\alpha\sqrt{n}}}{\sqrt{n}}\right)^{2} \cdot \frac{1}{n-\alpha\sqrt{n}} \cdot \mathbb{E}
\left[ \sum_{\ell=1}^{n-\alpha\sqrt{n}} (I_{\ell}-q') \right]^{2} \\
&= \lim_{n \to \infty} \left(1-\frac{\sqrt{n-\alpha\sqrt{n}}}{\sqrt{n}}\right)^{2} \cdot \frac{1}{n-\alpha\sqrt{n}} \cdot  \sum_{\ell=1}^{n-\alpha\sqrt{n}} \mathbb{E}
\left[(I_{\ell}-q')^{2} \right] \\
&= \lim_{n \to \infty} \left(1-\frac{\sqrt{n-\alpha\sqrt{n}}}{\sqrt{n}}\right)^{2} \cdot \mathbb{E} \left[(I_{1}-q')^{2} \right] \\
&= 0,
\end{align}
which proves that that $|\tilde{X}_{n}-X_{n}|$ converges in $L^{2}$ to 0, thus also in distribution. It then follows from \cite[Theorem 3.1]{Billingsley} that $X_{n}$ converges in distribution to $X \sim \calN(0,\sigma^{2})$.

Concerning the sequence $Z_{n}$, consider the following
\begin{align} 
\lim_{n \to \infty} \mathbb{E} [Z_{n}]
&= \lim_{n \to \infty} \mathbb{E} \left[\frac{1}{\sqrt{n}} \left(\sum_{m=1}^{\alpha \sqrt{n}-1} K_{m} + 2 \right) \right] \\
&= \lim_{n \to \infty} \frac{1}{\sqrt{n}} \left(\sum_{m=1}^{\alpha \sqrt{n}-1} q' + 2 \right)  \\
&= \alpha q',
\end{align}
and furthermore,
\begin{align}
\lim_{n \to \infty} \textbf{Var} [Z_{n}]
&= \lim_{n \to \infty} \textbf{Var} \left[\frac{1}{\sqrt{n}} \left(\sum_{m=1}^{\alpha \sqrt{n}-1} K_{m} + 2 \right) \right] \\
&= \lim_{n \to \infty} \frac{1}{n} \sum_{m=1}^{\alpha \sqrt{n}-1} q'(1-q')  \\
&= 0.
\end{align}
It follows that $Z_{n}$ converges in $L^{2}$ to $Z=\alpha q'$, i.e., a deterministic random variable. Hence, $Z_{n}$ also converges to $Z=\alpha q'$ in probability \cite[Lemma 1.3.5]{DURRETT}. Now, for $\epsilon > 0$ arbitrarily small, consider the following  
\begin{align}
P_{n}
&= \pr\left\{X_{n} \geq Y_{n}+Z_{n} + \alpha q' \right\} \\
\label{ToExp1}
&= \pr\left\{X_{n} \geq Y_{n}+Z_{n} + \alpha q' | Z_{n} \geq \alpha q' - \epsilon \right\} \pr\left\{Z_{n} \geq \alpha q' - \epsilon \right\} \nn \\
&~~+ \pr\left\{X_{n} \geq Y_{n}+Z_{n} + \alpha q' | Z_{n} < \alpha q' - \epsilon \right\} \pr\left\{Z_{n} < \alpha q' - \epsilon \right\} \\
&\leq \pr\left\{X_{n} \geq Y_{n}+ \alpha q' - \epsilon + \alpha q' | Z_{n} \geq \alpha q' - \epsilon \right\} \pr\left\{Z_{n} \geq \alpha q' - \epsilon \right\} +  \pr\left\{Z_{n} < \alpha q' - \epsilon \right\} \\
\label{ToExp2}
&= \pr\left\{X_{n} \geq Y_{n}+ 2\alpha q' - \epsilon \right\} \pr\left\{Z_{n} \geq \alpha q' - \epsilon \right\} +  \pr\left\{Z_{n} < \alpha q' - \epsilon \right\},
\end{align}
where \eqref{ToExp1} is due to the law of total probability and \eqref{ToExp2} follows from the fact that $(X_{n},Y_{n})$ are independent of $Z_{n}$.
Since $\{I_{\ell}\}$ and $\{J_{s}\}$ are all independent, the joint law of the pair $(X_{n},Y_{n})$ converges to the joint law of $(X,Y)$ and $X,Y$ are independent. Hence, by Portmanteau's theorem \cite[p.\ 16, Theorem 2.1]{Billingsley}, and the fact that $Z_{n}$ converges to $Z=\alpha q'$ in probability,   
\begin{align}
\limsup_{n \to \infty} P_{n}
&\leq \pr \left\{X - Y \geq 2\alpha q' - \epsilon \right\} \\
&= \pr \left\{\calN(0,2q(1-q)) \geq 2\alpha q' - \epsilon \right\} \\
&= Q\left(t_{0}^{-}(\epsilon)\right),
\end{align} 
where 
\begin{align}
t_{0}^{-}(\epsilon) = \frac{2\alpha q' - \epsilon}{\sqrt{2q(1-q)}},
\end{align}
and
\begin{align}
Q(t) \dfn \int_{t}^{\infty} \frac{1}{\sqrt{2\pi}} \exp\left\{-\frac{s^{2}}{2}\right\} ds.
\end{align}
In a similar fashion, 
\begin{align}
P_{n}
&= \pr\left\{X_{n} \geq Y_{n}+Z_{n} + \alpha q' \right\} \\
&= \pr\left\{X_{n} \geq Y_{n}+Z_{n} + \alpha q' | Z_{n} \leq \alpha q' + \epsilon \right\} \pr\left\{Z_{n} \leq \alpha q' + \epsilon \right\} \nn \\
&~~+ \pr\left\{X_{n} \geq Y_{n}+Z_{n} + \alpha q' | Z_{n} > \alpha q' - \epsilon \right\} \pr\left\{Z_{n} > \alpha q' - \epsilon \right\} \\
&\geq \pr\left\{X_{n} \geq Y_{n}+ \alpha q' + \epsilon + \alpha q' | Z_{n} \leq \alpha q' + \epsilon \right\} \pr\left\{Z_{n} \leq \alpha q' + \epsilon \right\} \\ 
&= \pr\left\{X_{n} \geq Y_{n}+ 2\alpha q' + \epsilon \right\} \pr\left\{Z_{n} \leq \alpha q' + \epsilon \right\},
\end{align} 
and thus,
\begin{align}
\liminf_{n \to \infty} P_{n}
&\geq \pr \left\{X - Y \geq 2\alpha q' + \epsilon \right\} \\
&= \pr \left\{\calN(0,2q(1-q)) \geq 2\alpha q' + \epsilon \right\} \\
&= Q\left(t_{0}^{+}(\epsilon)\right),
\end{align} 
where 
\begin{align}
t_{0}^{+}(\epsilon) = \frac{2\alpha q' + \epsilon}{\sqrt{2q(1-q)}}.
\end{align}
From the continuity of the $Q$-function and the fact that $\epsilon>0$ is arbitrarily small, we conclude that 
\begin{align}
\limsup_{n \to \infty} P_{n}
\leq Q\left(t_{0}\right) \leq \liminf_{n \to \infty} P_{n},
\end{align}
where 
\begin{align}
\label{ToRef13}
t_{0} = \sqrt{\frac{2\alpha^{2} (1-q)}{q}},
\end{align}
and hence,
\begin{align}
\lim_{n \to \infty} P_{n} = Q\left(t_{0}\right).
\end{align}
Now, for any $\alpha > 0$ and $q \in (0,1)$, the expression in \eqref{ToRef13} is strictly positive and thus $\lim_{n \to \infty} P_{n} = Q(t_{0}) < \tfrac{1}{2}$. We conclude that for all $0 < \delta < \tfrac{1}{2} - Q(t_{0})$, $P_{n} \leq Q(t_{0}) + \delta < \tfrac{1}{2}$ holds for all sufficiently large $n$.

\subsubsection*{Step 2: Many Zeros with High Probability}

Let $0 < \delta < \tfrac{1}{2} - Q(t_{0})$ be given.
Let $Q_{n}^{0}, Q_{n}^{1}$ denote the probabilities of deciding `0', for the two possible initial states. 
Since $P_{n} \leq Q(t_{0}) + \delta < \tfrac{1}{2}$ for all sufficiently large $n$, it follows that $\min\{Q_{n}^{0},Q_{n}^{1}\} \geq \Phi(t_{0}) - \delta > \tfrac{1}{2}$ for all sufficiently large $n$, where $\Phi(t)$ is defined in \eqref{DEF_PHI}. 

Let $\epsilon > \delta >0$ such that $\Phi(t_{0}) - \delta > \Phi(t_{0}) - \epsilon > \tfrac{1}{2}$. We now prove that the probability of drawing a relatively small number of zeros tends to 0 as $n \to \infty$. Denote $N_{0}=N(\bX_{1};0)$ and consider the following for $s \geq 0$
\begin{align}
\pr \left\{N_{0} \leq 2n (\Phi(t_{0})-\epsilon) \right\}
&= \pr \left\{e^{-sN_{0}} \geq e^{-2ns (\Phi(t_{0})-\epsilon)} \right\} \\
\label{B_ToExp3}
&\leq \frac{\mathbb{E} \left[e^{-sN_{0}}\right]}{e^{-2ns (\Phi(t_{0})-\epsilon)}},
\end{align}   
where \eqref{B_ToExp3} is due to Markov's inequality. Since \eqref{B_ToExp3} holds for every $s \geq 0$, it follows that
\begin{align}
\label{ToRef15}
\pr \left\{N_{0} \leq 2n (\Phi(t_{0})-\epsilon) \right\}
\leq \inf_{s \geq 0} \frac{\mathbb{E} \left[e^{-sN_{0}}\right]}{e^{-2ns (\Phi(t_{0})-\epsilon)}}.
\end{align}
Note that
\begin{align}
N_{0} = \sum_{\ell=1}^{n+\alpha\sqrt{n}} I_{\ell} + \sum_{k=1}^{n-\alpha\sqrt{n}} J_{k},
\end{align}
where $I_{\ell} \sim \text{Ber}(Q_{n}^{0})$, for all $\ell \in \{1,2,\ldots, n+\alpha\sqrt{n}\}$, $J_{k} \sim \text{Ber}(Q_{n}^{1})$, for all $k \in \{1,2,\ldots, n-\alpha\sqrt{n}\}$, and all of these binary random variables are independent.
We get that
\begin{align}
\mathbb{E}\left[e^{-s N_{0}}\right]
&= \mathbb{E}\left[\exp\left\{-s \left(\sum_{\ell=1}^{n+\alpha\sqrt{n}} I_{\ell} + \sum_{k=1}^{n-\alpha\sqrt{n}} J_{k}\right) \right\}\right] \\
&= \mathbb{E}\left[\prod_{\ell=1}^{n+\alpha\sqrt{n}} e^{-s I_{\ell}} \cdot \prod_{k=1}^{n-\alpha\sqrt{n}} e^{-s J_{k}} \right] \\
\label{B_ToExp4}
&= \prod_{\ell=1}^{n+\alpha\sqrt{n}} \mathbb{E}\left[ e^{-s I_{\ell}} \right] \cdot \prod_{k=1}^{n-\alpha\sqrt{n}} \mathbb{E}\left[ e^{-s J_{k}} \right] \\
&= \left[1+Q_{n}^{0}(e^{-s}-1)\right]^{n+\alpha\sqrt{n}} \cdot \left[1+Q_{n}^{1}(e^{-s}-1)\right]^{n-\alpha\sqrt{n}}  \\
\label{B_ToExp5}
&\leq \left[1+(\Phi(t_{0})-\delta)(e^{-s}-1)\right]^{n+\alpha\sqrt{n}} \cdot \left[1+(\Phi(t_{0})-\delta)(e^{-s}-1)\right]^{n-\alpha\sqrt{n}}  \\
\label{ToRef14}
&= \left[1+(\Phi(t_{0})-\delta)(e^{-s}-1)\right]^{2n},
\end{align}
where \eqref{B_ToExp4} is due to the independence of all binary random variables and \eqref{B_ToExp5} is true since $\min\{Q_{n}^{0},Q_{n}^{1}\} \geq \Phi(t_{0})-\delta$ for all sufficiently large $n$ and $e^{-s}-1 \leq 0$.  
Substituting \eqref{ToRef14} back into \eqref{ToRef15} yields that
\begin{align}
\pr \left\{N_{0} \leq 2n (\Phi(t_{0})-\epsilon) \right\}
&\leq \inf_{s \geq 0} \exp \left\{2n \log \left[1+(\Phi(t_{0})-\delta)(e^{-s}-1)\right] + 2ns (\Phi(t_{0}) - \epsilon)\right\} \\
\label{ToRef16}
&= \exp \left\{ 2n \cdot \inf_{s \geq 0} \{\log \left[1+(\Phi(t_{0})-\delta)(e^{-s}-1)\right] + s (\Phi(t_{0}) - \epsilon) \}\right\}.
\end{align}
Upon defining 
\begin{align}
\label{ToRef17}
g(s) = \log \left[1+(\Phi(t_{0})-\delta)(e^{-s}-1)\right] + s (\Phi(t_{0}) - \epsilon),
\end{align}
we find that the solution to $g'(s)=0$ is given by 
\begin{align}
s^{*} = \log \left(\frac{(\Phi(t_{0})-\delta)[1-(\Phi(t_{0})-\epsilon)]}{[1-(\Phi(t_{0})-\delta)](\Phi(t_{0})-\epsilon)}\right).
\end{align}
Substituting it back into \eqref{ToRef17} yields that
\begin{align}
g(s^{*}) 
&= \log\left(1+ (\Phi(t_{0})-\delta)\left[\frac{[1-(\Phi(t_{0})-\delta)](\Phi(t_{0})-\epsilon)}{(\Phi(t_{0})-\delta)[1-(\Phi(t_{0})-\epsilon)]}-1\right]\right) \nn \\
&~~~~~~~~~~~~~+ (\Phi(t_{0})-\epsilon) \log \left(\frac{(\Phi(t_{0})-\delta)[1-(\Phi(t_{0})-\epsilon)]}{[1-(\Phi(t_{0})-\delta)](\Phi(t_{0})-\epsilon)}\right) \\
&= \log \left(\frac{1-(\Phi(t_{0})-\delta)}{1-(\Phi(t_{0})-\epsilon)}\right) 
+(\Phi(t_{0})-\epsilon) \log \left(\frac{\Phi(t_{0})-\delta}{\Phi(t_{0})-\epsilon}\right) \nn \\
&~~~~~~~~~~~~~+ (\Phi(t_{0})-\epsilon) \log \left(\frac{1-(\Phi(t_{0})-\epsilon)}{1-(\Phi(t_{0})-\delta)}\right)  \\
&= - (\Phi(t_{0})-\epsilon) \log \left(\frac{\Phi(t_{0}) - \epsilon}{\Phi(t_{0})-\delta}\right)
- (1-(\Phi(t_{0})-\epsilon)) \log \left(\frac{1-(\Phi(t_{0}) - \epsilon)}{1-(\Phi(t_{0})-\delta)}\right) \\
\label{ToDoPinsker}
&= -D(\Phi(t_{0})-\epsilon \| \Phi(t_{0})-\delta).
\end{align}
We upper-bound the expression in \eqref{ToDoPinsker} using Pinsker's inequality \cite{C1967,Kullback}. Recall that the total variation distance between two probability distributions $P$ and $Q$ is defined by
\begin{align} \label{DEF_TVD}
|P-Q| = \frac{1}{2} \sum_{x \in \calX} |P(x)-Q(x)|,
\end{align}   
and the Kullback--Leibler divergence is defined by
\begin{align}
D(P\|Q) = \sum_{x \in \calX} P(x) \log \frac{P(x)}{Q(x)}.
\end{align}
Then, Pinsker's inequality asserts that  
\begin{align} \label{PINSKER}
D(P\|Q) \geq 2 |P-Q|^{2}.
\end{align}
Thus, we arrive at 
\begin{align}
\pr \left\{N_{0} \leq 2n (\Phi(t_{0})-\epsilon) \right\}
&\leq \exp \left\{ -2n D(\Phi(t_{0})-\epsilon \| \Phi(t_{0})-\delta) \right\} \\
&\leq \exp \left\{ -4n (\epsilon-\delta)^{2} \right\}.
\end{align}
Hence, we conclude that for all $n$ sufficiently large 
\begin{align}
\pr \left\{N_{0} \geq 2n (\Phi(t_{0})-\epsilon) \right\}
&\geq 1 - \exp \left\{ -4n (\epsilon-\delta)^{2} \right\},
\end{align} 
which converges to 1 as $n \to \infty$. 
Proposition \ref{PROP_2} is now proved.

\section*{Appendix C - Proof of Proposition \ref{PROP_3}}
\renewcommand{\theequation}{C.\arabic{equation}}
\setcounter{equation}{0}  

Denote $N_{0}=N(\bX_{1};0)$.  
Let $\{p_{n}\}$ denote the sequence of probabilities of the events that an agent with an initial value `0' updates its value to `0' after a single round of communication. 

\subsubsection*{Step 1: An Upper Bound on the PMF of the Binomial Distribution}
We start by upper-bounding the probability mass function (PMF) of the binomial random variable $X=\text{Bin}(n,p)$, which is given by
\begin{align}
\label{TermToCall2}
P_{X}(k) = \binom{n}{k} p^{k} (1-p)^{n-k},~~~k \in [0:n].
\end{align}
In order to upper-bound the binomial coefficient in \eqref{TermToCall2}, we invoke the following Stirling's bounds:
\begin{align}
\label{Stirling}
\sqrt{2 \pi n} \cdot n^{n} \cdot e^{-n} 
\leq n! 
\leq e \sqrt{n} \cdot n^{n} \cdot e^{-n}, 
\end{align} 
and get the following
\begin{align}
\binom{n}{k}
&= \frac{n!}{k! \cdot (n-k)!} \\
&\leq \frac{e \sqrt{n} \cdot n^{n} \cdot e^{-n}}{\sqrt{2\pi k} \cdot k^{k} \cdot e^{-k} \cdot \sqrt{2\pi (n-k)} \cdot (n-k)^{n-k} \cdot e^{-(n-k)}} \\
&= \frac{e \sqrt{n} \cdot n^{n}}{\sqrt{2\pi k} \cdot k^{k} \cdot \sqrt{2\pi (n-k)} \cdot (n-k)^{n-k}} \\
&= \frac{e}{2\pi} \sqrt{\frac{n}{k(n-k)}}
\frac{n^{k} \cdot n^{n-k}}{k^{k} \cdot (n-k)^{n-k}} \\
&= \frac{e}{2\pi} \sqrt{\frac{n}{k(n-k)}}
\exp \left\{ -k \log \left(\frac{k}{n}\right) 
-(n-k) \log \left(\frac{n-k}{n}\right) \right\} \\
\label{TermToCall3}
&= \frac{e}{2\pi} \sqrt{\frac{n}{k(n-k)}}
\exp \left\{ -n \left[\frac{k}{n} \log \left(\frac{k}{n}\right) 
+ \left(1-\frac{k}{n}\right) \log \left(1-\frac{k}{n}\right) \right] \right\}.
\end{align}
Substituting \eqref{TermToCall3} back into \eqref{TermToCall2} yields
\begin{align}
P_{X}(k) 
&\leq \frac{e}{2\pi} \sqrt{\frac{n}{k(n-k)}}
\exp \left\{ -n \left[\frac{k}{n} \log \left(\frac{k}{n}\right) 
+ \left(1-\frac{k}{n}\right) \log \left(1-\frac{k}{n}\right) \right] \right\} \cdot p^{k} (1-p)^{n-k} \\
&= \frac{e}{2\pi} \sqrt{\frac{n}{k(n-k)}}
\exp \left\{ -n \left[\frac{k}{n} \log \left(\frac{k}{n}\right) 
+ \left(1-\frac{k}{n}\right) \log \left(1-\frac{k}{n}\right) \right] \right\} \nn \\
&~~~~~~~~~~~~~~~~~~~~~~~~~ \times \exp \left\{ -n \left[\frac{k}{n} \log \left(\frac{1}{p}\right) 
+ \left(1-\frac{k}{n}\right) \log \left(\frac{1}{1-p}\right) \right] \right\} \\
&= \frac{e}{2\pi} \sqrt{\frac{n}{k(n-k)}}
\exp \left\{ -n \left[\frac{k}{n} \log \left(\frac{k/n}{p}\right) 
+ \left(1-\frac{k}{n}\right) \log \left(\frac{1-k/n}{1-p}\right) \right] \right\} \\
\label{ToRef3}
&= \frac{e}{2\pi} \sqrt{\frac{n}{k(n-k)}}
\exp \left\{ -n D\left(\frac{k}{n} \middle\| p \right) \right\} ,
\end{align}
where $D(\alpha \| \beta)$, for $\alpha,\beta \in [0,1]$, is defined in \eqref{DEF_Bin_DIVERGENCE}.

\subsubsection*{Step 2: The Limit of $\{p_{n}\}$ is $\tfrac{1}{2}$}

First, we show that $\{p_{n}\}$ is lower-bounded by $\tfrac{1}{2}$. For $q' = 1-q$, denote
\begin{align}
Z \sim \text{Bin}\left(n-1,q'\right),~~
X,Y \sim \text{Bin}\left(n,q'\right).
\end{align}
We have that
\begin{align}
p_{n} 
&= \pr\left\{Z+1 \geq Y \right\}  \\
\label{C_ToExp1}
&\geq \pr\left\{\text{Bin}\left(n-1,q'\right)+ \text{Bin}\left(1,q'\right) \geq Y \right\}  \\
&= \pr\left\{X \geq Y \right\},
\end{align}
where \eqref{C_ToExp1} is true since $\text{Bin}\left(1,q'\right) \leq 1$ with probability one. It follows by symmetry that 
\begin{align}
1 &= \pr\{X > Y\}+\pr\{X < Y\}+\pr\{X = Y\} \\
&= 2\pr\{X > Y\}+\pr\{X = Y\},
\end{align}
or,
\begin{align}
\pr\{X > Y\} = \frac{1}{2} - \frac{1}{2} \cdot \pr\{X = Y\},
\end{align}
which implies that 
\begin{align}
p_{n} 
&\geq \pr\{X \geq Y\} \\ 
&= \pr\{X > Y\} + \pr\{X = Y\} \\
&= \frac{1}{2} + \frac{1}{2} \cdot \pr\{X = Y\} \\ &\geq \frac{1}{2}.
\end{align}
Next, we upper-bound the sequence $\{p_{n}\}$.
Note that
\begin{align}
p_{n} 
&= \pr\left\{\text{Bin}\left(n-1,q'\right) + 1 \geq \text{Bin}\left(n,q'\right) \right\}  \\
&\leq \pr\left\{\text{Bin}\left(n,q'\right) + 1 \geq \text{Bin}\left(n,q'\right) \right\}  \\
&= \pr\left\{X + 1 \geq Y \right\}  \\
&= \pr\left\{X \geq Y \right\} + \pr\left\{X + 1 = Y \right\} \\
\label{ToRef0}
&= \frac{1}{2} + \frac{1}{2} \cdot \pr\{X = Y\} + \pr\left\{X + 1 = Y \right\}.
\end{align}
As for the last term in \eqref{ToRef0}, we have that
\begin{align}
\pr\left\{X + 1 = Y \right\}
&= \sum_{\ell=0}^{n-1} \pr\{X=\ell\} \cdot \pr\{Y=\ell+1\} \\
\label{ToExp0}
&\leq \sqrt{\sum_{\ell=0}^{n-1} \left(\pr\{X=\ell\}\right)^{2}}
\sqrt{\sum_{\ell=0}^{n-1} \left(\pr\{Y=\ell+1\}\right)^{2}} \\
&= \sqrt{\sum_{\ell=0}^{n-1} \left(\pr\{X=\ell\}\right)^{2}}
\sqrt{\sum_{\ell=1}^{n} \left(\pr\{Y=\ell\}\right)^{2}} \\
&\leq \sqrt{\sum_{\ell=0}^{n} \left(\pr\{X=\ell\}\right)^{2}}
\sqrt{\sum_{\ell=0}^{n} \left(\pr\{Y=\ell\}\right)^{2}} \\
&=\sum_{\ell=0}^{n} \left(\pr\{X=\ell\}\right)^{2} \\
&=\sum_{\ell=0}^{n} \pr\{X=\ell\} \cdot \pr\{Y=\ell\}\\
\label{ToRef1}
&= \pr\{X=Y\},
\end{align} 
where \eqref{ToExp0} follows from the Cauchy-Schwarz inequality. Substituting \eqref{ToRef1} back into \eqref{ToRef0} yields that
\begin{align}
\label{ToRef6}
p_{n} 
&\leq \frac{1}{2} + \frac{3}{2} \cdot \pr\{X = Y\}.
\end{align}
Now, consider the following:
\begin{align}
\pr\{X = Y\}
&=\sum_{\ell=0}^{n} \left(\pr\{X=\ell\}\right)^{2}\\
&=\sum_{\ell=0}^{n} \left[\binom{n}{\ell} (1-q)^{\ell} q^{n-\ell} \right]^{2} \\
&= \left[\binom{n}{0} (1-q)^{0} q^{n} \right]^{2}
+\sum_{\ell=1}^{n-1} \left[\binom{n}{\ell} (1-q)^{\ell} q^{n-\ell} \right]^{2} 
+ \left[\binom{n}{n} (1-q)^{n} q^{0} \right]^{2}\\
\label{ToRef2}
&= q^{2n} 
+\sum_{\ell=1}^{n-1} \left[\binom{n}{\ell} (1-q)^{\ell} q^{n-\ell} \right]^{2} + (1-q)^{2n}. 
\end{align}
As for the middle term in \eqref{ToRef2}, it follows from \eqref{ToRef3} that
\begin{align}
\sum_{\ell=1}^{n-1} \left[\binom{n}{\ell} (1-q)^{\ell} q^{n-\ell} \right]^{2}
\label{ToRef4}
&\leq \sum_{\ell=1}^{n-1} 
\left(\frac{e}{2\pi}\right)^{2} \frac{n}{\ell(n-\ell)}
\exp \left\{ -2n D\left(\frac{\ell}{n} \middle\| 1-q \right) \right\}.
\end{align}
In order to upper-bound \eqref{ToRef4}, 
let $\epsilon_{n} = 1/\sqrt[4]{n}$, for $n=1,2, \ldots$ and define the set of numbers
\begin{align}
\calN_{n} = \{n(1-q-\epsilon_{n}),n(1-q-\epsilon_{n})+1,\ldots,n(1-q),\ldots,n(1-q+\epsilon_{n})\},
\end{align}
whose cardinality is given by
\begin{align}
\label{Cardinality}
|\calN_{n}| = 2 n \epsilon_{n} +1.
\end{align}
Denote $\calM_{n} = \{1,2,\ldots,n-1\} \cap \calN_{n}^{\mbox{\tiny c}}$.
For any $\ell \in \calM_{n}$, it follows from Pinsker's inequality that
\begin{align}
D\left(\frac{\ell}{n} \middle\| 1-q \right)
&\geq D\left(1-q+\epsilon_{n} \middle\| 1-q \right)\\
\label{Pinsker_Result}
&\geq 2 \epsilon_{n}^{2}.
\end{align}
We now continue from \eqref{ToRef4} and arrive at
\begin{align}
&\sum_{\ell=1}^{n-1} 
\left(\frac{e}{2\pi}\right)^{2} \frac{n}{\ell(n-\ell)}
\exp \left\{ -2n D\left(\frac{\ell}{n} \middle\| 1-q \right) \right\} \nn \\
\label{C_ToExp2}
&~~~\leq 
\sum_{\ell \in \calM_{n}} 
\left(\frac{e}{2\pi}\right)^{2} \frac{n}{\ell(n-\ell)}
\exp \left\{ -4n \epsilon_{n}^{2} \right\}
+
\sum_{\ell \in \calN_{n}} 
\left(\frac{e}{2\pi}\right)^{2} \frac{n}{\ell(n-\ell)} \\
\label{C_ToExp4}
&~~~\leq 
\sum_{\ell \in \calM_{n}} 
\left(\frac{e}{2\pi}\right)^{2} \frac{n}{(n-1)}
\exp \left\{ -4n \epsilon_{n}^{2} \right\}
+
\sum_{\ell \in \calN_{n}} 
\left(\frac{e}{2\pi}\right)^{2} \frac{n}{n(1-q-\epsilon_{n})[n-n(1-q-\epsilon_{n})]} \\
\label{C_ToExp3}
&~~~\leq  
\left(\frac{e}{2\pi}\right)^{2} n
\exp \left\{ -4n \epsilon_{n}^{2} \right\}
+ 
\left(\frac{e}{2\pi}\right)^{2} \frac{2 n \epsilon_{n} +1}{n(q+\epsilon_{n})(1-q-\epsilon_{n})} \\
\label{ToRef5}
&~~~=  
\left(\frac{e}{2\pi}\right)^{2} n
\exp \left\{ -4 n^{1/2} \right\}
+ 
\left(\frac{e}{2\pi}\right)^{2} \frac{2 n^{3/4} +1}{n(q+n^{-1/4})(1-q-n^{-1/4})},
\end{align}
where \eqref{C_ToExp2} follows from \eqref{Pinsker_Result} and the fact that $D(\alpha\|\beta) \geq 0$ in general. 
The inequality in \eqref{C_ToExp4} is because of the following reasons. First, the minimizers of $\ell(n-\ell)$ in $\calM_{n}$ are $1$ or $n-1$. Second, the minimizer of $\ell(n-\ell)$ in $\calN_{n}$ is the endpoint of $\calN_{n}$ which is the most distant from $1/2$. For simplicity, we assumed without loss of generality that $q \in (1/2,1)$.
The passage to \eqref{C_ToExp3} is due to the fact that $|\calM_{n}| \leq n-1$ as well as \eqref{Cardinality} and in \eqref{ToRef5}, we substituted $\epsilon_{n} = 1/\sqrt[4]{n}$.   
Denote the expression in \eqref{ToRef5} by $G_{n}$ and notice that this expression converges to zero as $n \to \infty$.  
We substitute $G_{n}$ back into \eqref{ToRef2} and then into \eqref{ToRef6}. Since $\{p_{n}\}$ is lower-bounded by $\tfrac{1}{2}$, we conclude that
\begin{align}
\frac{1}{2} \leq p_{n} \leq \frac{1}{2} + \frac{3}{2} \cdot \left[q^{2n} 
+ G_{n} + (1-q)^{2n}\right].
\end{align}
Thus, $\{p_{n}\}$ converges to $\tfrac{1}{2}$ as long as $q \neq 0,1$.

\subsubsection*{Step 3: Asymptotic Behavior of the Number of Zeros}
We would like to prove that the random variable $|N_{0}-n|/\sqrt{n}$ is bounded away from zero with an overwhelmingly high probability at large $n$.
Note that
\begin{align}
N_{0} = \sum_{\ell=1}^{n} I_{n,\ell} + \sum_{\ell=1}^{n} J_{n,\ell}, 
\end{align}
where $I_{n,\ell} \sim \text{Ber}(p_{n})$ and $J_{n,\ell} \sim \text{Ber}(1-p_{n})$, for all $\ell \in \{1,2,\ldots, n\}$, and all of these binary random variables are independent.
Let $\epsilon > 0$ and $\delta(\epsilon) > 0$, that will be specified later on with the property that $\delta(\epsilon) \xrightarrow{\epsilon \to 0} 0$.
Consider the following
\begin{align}
\pr \left\{\left|\frac{N_{0}-n}{\sqrt{n}}\right| \geq \delta(\epsilon) \right\}
\label{ToRef18}
= \pr \left\{\left|\frac{1}{\sqrt{n}}\sum_{\ell=1}^{n} (I_{n,\ell}-p_{n}) + \frac{1}{\sqrt{n}}\sum_{\ell=1}^{n} (J_{n,\ell}-(1-p_{n}))\right| \geq \delta(\epsilon) \right\}.
\end{align}
In order to conclude that the two normalized sums inside the probability in \eqref{ToRef18} converge in distribution to normal random variables, we invoke Lindeberg-Feller central limit theorem \cite[p.\ 116, Theorem 2.4.5.]{DURRETT}. First, we introduce the concept of a ``triangular array'' of variables. A triangular array of random variables is of the form $\{X_{n,i}\}$, $n \geq 1$, $1 \leq i \leq n$, where for every $n$, the random variables $X_{n,1}, X_{n,2}, \ldots, X_{n,n}$ are independent, have zero mean, and have finite variance. Then, one have the following result.
\begin{theorem}[Lindeberg-Feller CLT]
	Suppose $\{X_{n,i}\}$ is a triangular array such that
	\begin{align}
	Z_{n} &= \frac{1}{n}\sum_{i=1}^{n} X_{n,i}, \\
	s_{n}^{2} &= \frac{1}{n}\sum_{i=1}^{n} \textbf{Var} [X_{n,i}],
	\end{align}
	and $s_{n}^{2} \to s^{2} \neq 0$. If the Lindeberg condition holds: for every $\epsilon > 0$,
	\begin{align}
	\label{Lindeberg_condition}
	\frac{1}{n} \sum_{i=1}^{n} \mathbb{E} \left[X_{n,i}^{2} \IND\{|X_{n,i}| \geq \epsilon \sqrt{n}\}\right] \to 0,
	\end{align}
	then $\sqrt{n}Z_{n} \xrightarrow{d} \calN(0,s^{2})$.
\end{theorem}    
Now, concerning the left-hand-side normalized sum inside the probability in \eqref{ToRef18}, notice that
\begin{align}
s_{n}^{2} 
&= \frac{1}{n}\sum_{\ell=1}^{n} \textbf{Var} [I_{n,\ell}-p_{n}] \\
&= \frac{1}{n}\sum_{\ell=1}^{n} p_{n}(1-p_{n}) \\
&= p_{n}(1-p_{n}),
\end{align}
which converges to $s^{2} = \tfrac{1}{4}$ as $n \to \infty$. In addition, Lindeberg's condition in \eqref{Lindeberg_condition} is trivially satisfied since all the random variables in our setting are bounded. Thus, it follows by Lindeberg-Feller CLT that
\begin{align}
\frac{1}{\sqrt{n}}\sum_{\ell=1}^{n} (I_{n,\ell}-p_{n})
\xrightarrow{d} X \sim \calN(0,\tfrac{1}{4}).
\end{align}   
From exactly the same considerations,
\begin{align}
\frac{1}{\sqrt{n}}\sum_{\ell=1}^{n} (J_{n,\ell}-(1-p_{n}))
\xrightarrow{d} Y \sim \calN(0,\tfrac{1}{4}),
\end{align}
and $X,Y$ are independent since $\{I_{n,\ell}\}$ and $\{J_{n,\ell}\}$ are all independent.
We continue from \eqref{ToRef18} and arrive at
\begin{align}
\lim_{n \to \infty} \pr \left\{\left|\frac{N_{0}-n}{\sqrt{n}}\right| \geq \delta(\epsilon) \right\}
&= \pr \left\{\left|X + Y\right| \geq \delta(\epsilon) \right\} \\
&= \pr \left\{\left|\calN(0,\tfrac{1}{2})\right| \geq \delta(\epsilon) \right\} \\
&= 1- \frac{\epsilon}{2},
\end{align} 
which can obviously be satisfied by a proper choice of $\delta(\epsilon)$.
We conclude that for any $\epsilon > 0$, there exists some $M(\epsilon)$, such that for all $n \geq M(\epsilon)$, 
\begin{align}
\pr \left\{\left|\frac{N_{0}-n}{\sqrt{n}}\right| \geq \delta(\epsilon) \right\}
\geq 1- \epsilon,
\end{align} 
which completes the proof of Proposition \ref{PROP_3}.

\section*{Appendix D - Proof of Proposition \ref{PROP_4}}
\renewcommand{\theequation}{D.\arabic{equation}}
\setcounter{equation}{0} 

Let us denote $N=N(\bX_{1};0)$.
For any $\mu \geq 0$, it follows from Markov's inequality that
\begin{align}
\pr \left\{N \geq n + B_{n} \right\}
&= \pr \left\{e^{\mu N} \geq e^{\mu(n + B_{n})} \right\} \\
\label{D_ToRef1}
&\leq \frac{\mathbb{E}\left[e^{\mu N}\right]}{e^{\mu(n + B_{n})}},
\end{align}
and thus, since \eqref{D_ToRef1} holds for every $\mu \geq 0$, it follows that
\begin{align}
\label{ToRef9}
\pr \left\{N \geq n + B_{n} \right\}
&\leq \inf_{\mu \geq 0} \frac{\mathbb{E}\left[e^{\mu N}\right]}{e^{\mu(n + B_{n})}}.
\end{align}
Note that
\begin{align}
N = \sum_{m=1}^{n} I_{m} + \sum_{m=1}^{n} J_{m},
\end{align}
where $I_{m} \sim \text{Ber}(p_{n})$ and $J_{m} \sim \text{Ber}(1-p_{n})$, for all $m \in \{1,2,\ldots, n\}$, and all of these binary random variables are independent.
We get that
\begin{align}
\mathbb{E}\left[e^{\mu N}\right]
&= \mathbb{E}\left[\exp\left\{\mu \left(\sum_{m=1}^{n} I_{m} + \sum_{m=1}^{n} J_{m}\right) \right\}\right] \\
&= \mathbb{E}\left[\prod_{m=1}^{n} e^{\mu I_{m}} \cdot \prod_{m=1}^{n} e^{\mu J_{m}} \right] \\
\label{D_ToRef2}
&= \prod_{m=1}^{n} \mathbb{E}\left[ e^{\mu I_{m}} \right] \cdot \prod_{m=1}^{n} \mathbb{E}\left[ e^{\mu J_{m}} \right] \\
&= \left(1-p_{n} + p_{n}e^{\mu}\right)^{n} \cdot \left(p_{n} + (1-p_{n})e^{\mu}\right)^{n} \\
\label{D_ToRef3}
&= \left\{\left[1+p_{n}(e^{\mu}-1)\right] \cdot \left[1+(1-p_{n})(e^{\mu}-1)\right] \right\}^{n} \\
\label{D_ToRef5}
&\leq \left\{\left[1+\tfrac{1}{2}(e^{\mu}-1)\right] \cdot \left[1+\tfrac{1}{2}(e^{\mu}-1)\right] \right\}^{n} \\ 
\label{D_ToRef6}
&= \left[1+\tfrac{1}{2}(e^{\mu}-1)\right]^{2n}.
\end{align}
where \eqref{D_ToRef2} is due to the independence of all binary random variables and \eqref{D_ToRef5} follows from the fact that the expression in \eqref{D_ToRef3} is maximized for $p_{n}=\tfrac{1}{2}$.

Substituting \eqref{D_ToRef6} back into \eqref{ToRef9} yields that
\begin{align}
\pr \left\{N \geq n + B_{n} \right\}
&\leq \inf_{\mu \geq 0} \frac{\left[1+\tfrac{1}{2}(e^{\mu}-1)\right]^{2n}}{\exp\{\mu(n + B_{n})\}} \\
&= \inf_{\mu \geq 0} \exp \left\{2n \log \left[1+\tfrac{1}{2}(e^{\mu}-1)\right] - \mu(n + B_{n})\right\} \\
\label{ToRef10}
&= \exp \left\{\inf_{\mu \geq 0} \{2n \log \left[1+\tfrac{1}{2}(e^{\mu}-1)\right]-\mu(n + B_{n}) \} \right\}.
\end{align}
Upon defining 
\begin{align}
f(\mu) = 2n \log \left[1+\tfrac{1}{2}(e^{\mu}-1)\right]-\mu(n + B_{n}),
\end{align}
we find that the solution to $f'(\mu)=0$ is given by 
\begin{align}
\mu^{*}
= \log \left(\frac{n+B_{n}}{n-B_{n}}\right).
\end{align}
Substituting it back into \eqref{ToRef10} provides that
\begin{align}
&\pr \left\{N \geq n + B_{n} \right\} \nn \\
&\leq \exp \left\{2n \log \left[1+\tfrac{1}{2}(e^{\mu^{*}}-1)\right]-\mu^{*}(n + B_{n}) \right\} \\
&= \exp \left\{2n \log \left[1+\frac{1}{2}\left(\frac{n+B_{n}}{n-B_{n}}-1\right)\right] - (n + B_{n}) \log \left(\frac{n+B_{n}}{n-B_{n}}\right) \right\} \\
&= \exp \left\{2n \log \left(1 + \frac{B_{n}}{n-B_{n}} \right) - (n + B_{n}) \log \left(\frac{n+B_{n}}{n-B_{n}}\right) \right\} \\
&= \exp \left\{2n \log \left(\frac{n}{n-B_{n}} \right) - (n + B_{n}) \log \left(\frac{n+B_{n}}{n-B_{n}}\right) \right\} \\
&= \exp \left\{(n-B_{n}) \log \left(\frac{n}{n-B_{n}} \right) + (n+B_{n}) \log \left(\frac{n}{n+B_{n}} \right) \right\} \\
&= \exp \left\{-n \cdot \left[\left(1-\frac{B_{n}}{n}\right) \log \left(1-\frac{B_{n}}{n}\right) + \left(1+\frac{B_{n}}{n}\right) \log \left(1+\frac{B_{n}}{n}\right)\right] \right\}.
\end{align} 
Consider the function
\begin{align}
g(t) = \left(1-t\right) \log \left(1-t\right) + \left(1+t\right) \log \left(1+t\right),
\end{align}
which is symmetric around $t=0$. Its first order and second order derivatives are given by
\begin{align}
g'(t) = \log\left(\frac{1+t}{1-t}\right), 
\end{align}
and 
\begin{align}
g''(t) = \frac{2}{(1+t)(1-t)}.
\end{align}
Hence, we conclude that $g(t) \geq t^{2}$, and thus
\begin{align}
\pr \left\{N \geq n + B_{n} \right\}
\leq \exp \left\{-n \cdot \left(\frac{B_{n}}{n}\right)^{2} \right\}
= \exp \left\{-\frac{B_{n}^{2}}{n} \right\},
\end{align}
which completes the proof of Proposition \ref{PROP_4}.

\section*{Appendix E - Proof of Proposition 5}
\renewcommand{\theequation}{E.\arabic{equation}}
\setcounter{equation}{0}  

\subsubsection*{Step 1: A Simplification for the Consensus Probability}
Due to symmetry, we only analyze the case $\mathsf{I}_{0} > \mathsf{I}_{1}$. 
It follows that 
\begin{align}
\pr\{\calC_{n}\} 
&= \pr\{N(\bX_{1};0) = 2n\}\\
&= \pr \left\{\bigcap_{i=1}^{2n} \{\bX_{1}(i) = 0\} \right\} \\
&= \prod_{i=1}^{2n} \pr \left\{ \bX_{1}(i) = 0 \right\} \\
\label{ToSubs2}
&= \prod_{i=1}^{2n} \left(1-\pr\left\{ \bX_{1}(i) = 1 \right\}\right).
\end{align}

\subsubsection*{Step 2: A Lower Bound on $\pr\left\{ \bX_{1}(i) = 1 \right\}$}
If an agent starts with a `0', then the probability to decide in favor of `1' is lower-bounded by 
\begin{align}
&\pr\left\{\text{Bin}\left(n-C_{n},q'\right) \geq \text{Bin}\left(n+C_{n}-1,q'\right) +2 \right\} \nn \\
\label{ToCallE0}
&~~\geq \pr\left\{\text{Bin}\left(n-C_{n},q'\right) \geq \text{Bin}\left(n+C_{n},q'\right) + 2 \right\}.
\end{align} 
If an agent starts with a `1', then the probability to decide in favor of `1' is lower-bounded by 
\begin{align}
&\pr\left\{\text{Bin}\left(n-C_{n}-1,q'\right) + 1 \geq \text{Bin}\left(n+C_{n},q'\right) \right\} \nn \\
&~~\geq \pr\left\{\text{Bin}\left(n-C_{n}-1,q'\right) + \text{Bin}\left(1,q'\right) \geq \text{Bin}\left(n+C_{n},q'\right) \right\}  \\
\label{ToCallE1}
&~~= \pr\left\{\text{Bin}\left(n-C_{n},q'\right) \geq \text{Bin}\left(n+C_{n},q'\right) \right\}.
\end{align}
Since \eqref{ToCallE0} cannot be larger than \eqref{ToCallE1}, we continue with \eqref{ToCallE0}.
From now on, we lower-bound the probability in \eqref{ToCallE0}, to be denoted by $Q_{n}$.
The probability in \eqref{ToCallE0} can be written explicitly as 
\begin{align}
\label{ToCont0}
Q_{n} = \sum_{\ell=0}^{n-C_{n}} \sum_{k=0}^{n+C_{n}}
\binom{n-C_{n}}{\ell} (1-q)^{\ell} q^{n-C_{n}-\ell}
\binom{n+C_{n}}{k} (1-q)^{k} q^{n+C_{n}-k} \IND \{\ell \geq k+2\}.
\end{align} 
We continue by lower-bounding the PMF of the binomial random variable $X=\text{Bin}(n,p)$, which is given by
\begin{align}
\label{TermToCall4}
P_{X}(k) = \binom{n}{k} p^{k} (1-p)^{n-k},~~~k \in [0:n].
\end{align}
In order to lower-bound the binomial coefficient in \eqref{TermToCall4}, we use the Stirling's bounds in \eqref{Stirling} and get that
\begin{align}
\binom{n}{k}
&= \frac{n!}{k! \cdot (n-k)!} \\
\label{TermToCall5}
&\geq \frac{\sqrt{2\pi}}{e^{2}} \sqrt{\frac{n}{k(n-k)}}
\exp \left\{ -n \left[\frac{k}{n} \log \left(\frac{k}{n}\right) 
+ \left(1-\frac{k}{n}\right) \log \left(1-\frac{k}{n}\right) \right] \right\}.
\end{align}
Substituting \eqref{TermToCall5} back into \eqref{TermToCall4} yields
\begin{align}
P_{X}(k) 
\label{ToRef7}
&\geq \frac{\sqrt{2\pi}}{e^{2}} \sqrt{\frac{n}{k(n-k)}}
\exp \left\{ -n D\left(\frac{k}{n} \middle\| p \right) \right\} ,
\end{align}
where $D(\alpha \| \beta)$, for $\alpha,\beta \in [0,1]$, is defined in \eqref{DEF_Bin_DIVERGENCE}.  
Substituting twice this lower bound into \eqref{ToCont0}, we arrive at  
\begin{align}
\label{ToCont1}
Q_{n} 
&\geq 
\frac{2\pi}{e^{4}}
\sum_{\ell=0}^{n-C_{n}} \sum_{k=0}^{\ell-2}
\sqrt{\frac{n-C_{n}}{\ell(n-C_{n}-\ell)}}
\exp \left\{ -(n-C_{n}) D\left(\frac{\ell}{n-C_{n}} \middle\| 1-q \right) \right\} \nn \\
&~~~~~~~~~~~~~~~~~~~~~\times 
\sqrt{\frac{n+C_{n}}{k(n+C_{n}-k)}}
\exp \left\{ -(n+C_{n}) D\left(\frac{k}{n+C_{n}} \middle\| 1-q \right) \right\}.
\end{align}
As for the square-root factors in \eqref{ToCont1}, we have the following
\begin{align}
&\sqrt{\frac{n-C_{n}}{\ell(n-C_{n}-\ell)}} \cdot
\sqrt{\frac{n+C_{n}}{k(n+C_{n}-k)}} \nn \\
\label{ToExp3}
&~~\geq \sqrt{\frac{n-C_{n}}{\tfrac{1}{2}(n-C_{n})(n-C_{n}-\tfrac{1}{2}(n-C_{n}))}} \cdot
\sqrt{\frac{n+C_{n}}{\tfrac{1}{2}(n+C_{n})(n+C_{n}-\tfrac{1}{2}(n+C_{n}))}} \\
&~~= \sqrt{\frac{4(n-C_{n})}{(n-C_{n})^{2}}} \cdot
\sqrt{\frac{4(n+C_{n})}{(n+C_{n})^{2}}} \\
&~~= 4\sqrt{\frac{1}{(n-C_{n})(n+C_{n})}} \\
&~~= 4\sqrt{\frac{1}{n^{2}-C_{n}^{2}}} \\
\label{ToCont2}
&~~\geq \frac{4}{n},
\end{align}
where \eqref{ToExp3} is due to the fact that a square has the maximal area among all rectangles with a fixed perimeter. 
Lower-bounding \eqref{ToCont1} using \eqref{ToCont2} yields 
\begin{align}
Q_{n} 
&\geq 
\frac{8\pi}{e^{4}n}
\sum_{\ell=0}^{n-C_{n}} \sum_{k=0}^{\ell-2}
\exp \left\{ -(n-C_{n}) D\left(\frac{\ell}{n-C_{n}} \middle\| 1-q \right) \right\} \nn \\
&~~~~~~~~~~~~~~~~~~~~~~~~~\times 
\exp \left\{ -(n+C_{n}) D\left(\frac{k}{n+C_{n}} \middle\| 1-q \right) \right\} \\
\label{ToExp7}
&\geq 
\frac{8\pi}{e^{4}n}
\sum_{\ell=(n-C_{n})(1-q)}^{(n+C_{n})(1-q)}
\sum_{k=0}^{\ell-2}
\exp \left\{ -(n-C_{n}) D\left(\frac{\ell}{n-C_{n}} \middle\| 1-q \right) \right\} \nn \\
&~~~~~~~~~~~~~~~~~~~~~~~~~~~~~~~~~~~~~~~~~~\times 
\exp \left\{ -(n+C_{n}) D\left(\frac{k}{n+C_{n}} \middle\| 1-q \right) \right\} \\
\label{ToExp8}
&\geq 
\frac{8\pi}{e^{4}n}
\sum_{\ell=(n-C_{n})(1-q)}^{(n+C_{n})(1-q)}
\sum_{k=\ell - C_{n}-2}^{\ell-2}
\exp \left\{ -(n-C_{n}) D\left(\frac{\ell}{n-C_{n}} \middle\| 1-q \right) \right\} \nn \\
&~~~~~~~~~~~~~~~~~~~~~~~~~~~~~~~~~~~~~~~~~~\times 
\exp \left\{ -(n+C_{n}) D\left(\frac{k}{n+C_{n}} \middle\| 1-q \right) \right\} \\
\label{ToExp9}
&=
\frac{8\pi}{e^{4}n}
\sum_{\ell=(n-C_{n})(1-q)}^{(n+C_{n})(1-q)}
\sum_{j=0}^{C_{n}}
\exp \left\{ -(n-C_{n}) D\left(\frac{\ell}{n-C_{n}} \middle\| 1-q \right) \right\} \nn \\
&~~~~~~~~~~~~~~~~~~~~~~~~~~~~~~~~~~~~~~~~~~\times 
\exp \left\{ -(n+C_{n}) D\left(\frac{\ell-2-j}{n+C_{n}} \middle\| 1-q \right) \right\} \\
\label{ToCont3}
&=
\frac{8\pi}{e^{4}n}
\sum_{m=0}^{2C_{n}} \sum_{j=0}^{C_{n}}
\exp \left\{ -(n-C_{n}) D\left(\frac{(n-C_{n}+m)(1-q)}{n-C_{n}} \middle\| 1-q \right) \right\} \nn \\
&~~~~~~~~~~~~~~~~~~~~~~~~~\times 
\exp \left\{ -(n+C_{n}) D\left(\frac{(n-C_{n}+m)(1-q)-2-j}{n+C_{n}} \middle\| 1-q \right) \right\},
\end{align}
where \eqref{ToExp7} follows from the condition $\lim_{n \to \infty} C_{n}/n = 0$, which implies that for all large enough $n$, both $(n-C_{n})(1-q) \geq 0$ and $(n+C_{n})(1-q) \leq n-C_{n}$ hold.
The inequality in \eqref{ToExp8} also follows from the condition $\lim_{n \to \infty} C_{n}/n = 0$, since for all $(n-C_{n})(1-q) \leq \ell \leq (n+C_{n})(1-q)$, it holds that $\ell - C_{n} - 2 \geq 0$, for all sufficiently large $n$.
In \eqref{ToExp9} we changed the summation index from $k$ to $j$ according to $k=\ell-j-2$, with $j \in \{0,1,\ldots,C_{n}\}$, and in \eqref{ToCont3} we changed the summation index from $\ell$ to $m$ according to $\ell=(n-C_{n}+m)(1-q)$, with $m \in \{0,1,\ldots,2C_{n}\}$. 
In order to upper-bound the divergence terms in \eqref{ToCont3}, we invoke the following reverse Pinsker inequality \cite[p.\ 5974, Eq.\ (23)]{SASON}
\begin{align} \label{Reverse_PINSKER}
D(P\|Q) \leq \left(\frac{2}{Q_{\mbox{\tiny min}}}\right) \cdot |P-Q|^{2},
\end{align}  
when
\begin{align}
Q_{\mbox{\tiny min}} = \min_{x\in \calX} Q(x).
\end{align}
Let us define $\Delta_{q}=\min\{q,1-q\}$.
Then, after some algebraic work, we arrive at
\begin{align}
Q_{n}
&\geq
\frac{8\pi}{e^{4}n}
\sum_{m=0}^{2C_{n}} \sum_{j=0}^{C_{n}}
\exp \left\{ -(n-C_{n}) \cdot \frac{2}{\Delta_{q}}  \frac{(1-q)^{2} m^{2}}{(n-C_{n})^{2}} \right\} \nn \\
&~~~~~~~~~~~~~~~~~~~~~~~~~\times 
\exp \left\{ -(n+C_{n}) \cdot \frac{2}{\Delta_{q}} \frac{[(1-q)(2C_{n}-m)+2+j]^{2}}{(n+C_{n})^{2}} \right\} \\
&=
\frac{8\pi}{e^{4}n}
\sum_{m=0}^{2C_{n}} \sum_{j=0}^{C_{n}}
\exp \left\{ - \frac{2}{\Delta_{q}}  \frac{(1-q)^{2} m^{2}}{n-C_{n}} \right\} \cdot
\exp \left\{ - \frac{2}{\Delta_{q}} \frac{[(1-q)(2C_{n}-m)+2+j]^{2}}{n+C_{n}} \right\} \\
\label{ToExp10}
&\geq
\frac{8\pi}{e^{4}n}
\sum_{m=0}^{2C_{n}} \sum_{j=0}^{C_{n}}
\exp \left\{ - \frac{2}{\Delta_{q}}  \frac{m^{2}}{n-C_{n}} \right\} \cdot
\exp \left\{ - \frac{2}{\Delta_{q}} \frac{[(2C_{n}-m)+2C_{n}]^{2}}{n-C_{n}} \right\} \\
&\geq
\frac{8\pi C_{n}}{e^{4}n}
\sum_{m=0}^{2C_{n}}
\exp \left\{ - \frac{2}{\Delta_{q}}  \frac{m^{2}}{n-C_{n}} \right\} \cdot
\exp \left\{ - \frac{2}{\Delta_{q}} \frac{(4C_{n}-m)^{2}}{n-C_{n}} \right\} \\
&=
\frac{8\pi C_{n}}{e^{4}n}
\sum_{m=0}^{2C_{n}}
\exp \left\{ - \frac{2}{\Delta_{q}} \cdot  \frac{m^{2}+(4C_{n}-m)^{2}}{n-C_{n}} \right\} \\
\label{ToRef20}
&=
\frac{8\pi C_{n}}{e^{4}n}
\sum_{m=0}^{2C_{n}}
\exp \left\{ - \frac{4}{\Delta_{q}} \cdot  \frac{(2C_{n}-m)^{2} + 4C_{n}^{2}}{n-C_{n}} \right\},
\end{align}
where \eqref{ToExp10} is true since $1 \geq 1-q$, $n-C_{n} \leq n+C_{n}$, and due to the fact that $2+j$ is obviously upper-bounded by $2C_{n}$. 
Now, the exponent in \eqref{ToRef20} is maximized at $m=0$, and thus
\begin{align}
Q_{n} 
&\geq \frac{8\pi C_{n}}{e^{4}n}
\sum_{m=0}^{2C_{n}}
\exp \left\{ - \frac{4}{\Delta_{q}} \cdot  \frac{8C_{n}^{2}}{n-C_{n}} \right\} \\
&\geq \frac{16\pi}{e^{4}} \frac{C_{n}^{2}}{n}
\exp \left\{-\frac{32}{\Delta_{q}} \cdot  \frac{C_{n}^{2}}{n-C_{n}} \right\}. 
\end{align}

\subsubsection*{Step 3: Wrapping Up}
We denote the constant $f_{q}=32/\Delta_{q}$.
Continuing from \eqref{ToSubs2}, we finally arrive at
\begin{align}
\pr\{\calC_{n}\} 
&\leq \prod_{i=1}^{2n} \left(1-\frac{16\pi}{e^{4}} \frac{C_{n}^{2}}{n} \cdot
\exp \left\{- f_{q} \cdot  \frac{C_{n}^{2}}{n-C_{n}} \right\}\right) \\
&= \left(1-\frac{16\pi}{e^{4}} \frac{C_{n}^{2}}{n} \cdot
\exp \left\{- f_{q} \cdot  \frac{C_{n}^{2}}{n-C_{n}} \right\}\right)^{2n} \\
&= \exp\left\{ 2n \cdot \log\left(1-\frac{16\pi}{e^{4}} \frac{C_{n}^{2}}{n} \cdot
\exp \left\{- f_{q} \cdot  \frac{C_{n}^{2}}{n-C_{n}} \right\}\right) \right\} \\
\label{ToExp6}
&\leq \exp\left\{- \frac{32\pi}{e^{4}} C_{n}^{2} \cdot
\exp \left\{- f_{q} \cdot  \frac{C_{n}^{2}}{n-C_{n}} \right\} \right\} \\
&\leq \exp\left\{- C_{n}^{2} \cdot
\exp \left\{- f_{q} \cdot  \frac{C_{n}^{2}}{n-C_{n}} \right\} \right\}, 
\end{align}
where \eqref{ToExp6} follows from the inequality $\log(1-y) \leq -y$.
This completes the proof of Proposition \ref{PROP_5}.


\end{document}